\documentclass[a4paper,11pt]{article}

\usepackage{jheppub} % for details on the use of the package, please
\usepackage[T1]{fontenc} % if needed

\usepackage{pgfplots}
\pgfplotsset{compat=newest}
\usepgfplotslibrary{fillbetween}

\usepackage{amssymb,amsmath}
\usepackage{bbm}
\usepackage{mathtools}
\usepackage{graphicx}
\usepackage{epsfig}
\usepackage{epstopdf}
\usepackage{comment}
\usepackage{setspace}
\usepackage{dsfont}
\usepackage{latexsym}
\usepackage{tikz}
\usepackage{psfrag}
\usepackage{booktabs}
\usepackage{bbm}
\usepackage[toc]{appendix}
\usepackage{color}
\usepackage{datetime}
\usepackage{tensor}
\usepackage{lscape}
\usepackage{lipsum}
\usepackage{caption}
\usepackage{subcaption}

\usepackage{tikz}
\usetikzlibrary{arrows,shapes}
\usetikzlibrary{trees}
\usetikzlibrary{patterns}
\usetikzlibrary{matrix,arrows} 				% For commutative diagram
\usetikzlibrary{positioning}				% For "above of=" commands
\usetikzlibrary{calc,through}				% For coordinates
\usetikzlibrary{decorations.pathreplacing}  % For curly braces
\usepackage{pgffor}							% For repeating patterns
\usetikzlibrary{decorations.pathmorphing}	% For Feynman Diagrams
\usetikzlibrary{decorations.markings}
\tikzset{
    vector/.style={decorate, decoration={snake}, draw},
	provector/.style={decorate, decoration={snake,amplitude=2.5pt}, draw},
	antivector/.style={decorate, decoration={snake,amplitude=-2.5pt}, draw},
        smallvector/.style={decorate, decoration={snake,amplitude=1.5pt,post length=0.5mm}, draw},
    fermion/.style={draw=black, postaction={decorate},
        decoration={markings,mark=at position .55 with {\arrow[draw=black]{>}}}},
    fermionbar/.style={draw=black, postaction={decorate},
        decoration={markings,mark=at position .55 with {\arrow[draw=black]{<}}}},
    fermionnoarrow/.style={draw=black},
    gluon/.style={decorate, draw=black,
        decoration={coil,amplitude=4pt, segment length=5pt}},
    scalar/.style={dashed,draw=black, postaction={decorate},
        decoration={markings,mark=at position .55 with {\arrow[draw=black]{>}}}},
    scalarbar/.style={dashed,draw=black, postaction={decorate},
        decoration={markings,mark=at position .55 with {\arrow[draw=black]{<}}}},
    scalarnoarrow/.style={dashed,draw=black},
    electron/.style={draw=black, postaction={decorate},
        decoration={markings,mark=at position .55 with {\arrow[draw=black]{>}}}},
    bigvector/.style={decorate, decoration={snake,amplitude=4pt}, draw},
    arrow/.style={draw=black, postaction={decorate},
        decoration={markings,mark=at position 1 with {\arrow[draw=black]{>}}}},
}

\tikzstyle{block} = [draw, rectangle, 
    minimum height=3em, minimum width=6earticlem]

\definecolor{darkblue}{rgb}{0.2, 0, 0.8}

\numberwithin{equation}{section}

\def\be{\begin{equation}}
\def\ee{\end{equation}}

\newcommand{\wt}{\widetilde}
\renewcommand{\d}{\delta}

\renewcommand{\a}{\alpha}

\newcommand{\eps}{\epsilon}

\newcommand{\Nd}{\mathcal{N}}

\newcommand{\Jd}{\mathcal{J}}

\newcommand{\Jt}{\ti{J}_1}

\newcommand{\gap}{\text{gap}}
\newcommand{\lop}{\text{loop}}
\newcommand{\sub}{\text{sub}}

\newcommand{\ti}{\Tilde}

\newcommand{\la}{\langle}
\newcommand{\ra}{\rangle}

\newcommand{\reef}[1]{(\ref{#1})}

\usepackage{xcolor}
\usepackage{pagecolor}

\begin{document}
\title{Analytic Bounds on the Spectrum of Crossing Symmetric S-Matrices}
\author{Justin Berman}
\affiliation{Leinweber Center for Theoretical Physics, Randall Laboratory of Physics,\\  
The University of Michigan, \\
450 Church Street, Ann Arbor, MI 48109-1040, USA}

\emailAdd{jdhb@umich.edu}

\abstract{We derive two rigorous constraints on the spectrum of massive states in weakly coupled theories with massless scalars in the adjoint representation of a large-$N$ gauge group. First, we show that the presence of massive spinning states necessitates the existence of lighter states with lower spins. Explicitly, if there exists a massive state with spin $J > 2$, then there must be a state with spin $J-1$ and a non-zero mass lower than that of the lightest spin-$J$ state, a state with spin $J-2$ and a mass lower than that of the lightest spin-($J-1$) particle and so on until we reach a mass below which only states with spin less than 2 are exchanged. Second, we find strict upper bounds on the masses of the lightest states at any spin. If there are spin-$J$ states in the spectrum, the maximum mass of the lightest spin-($J+1$) state is determined by the masses of the lightest spin-$J$ and $(J-1)$ states. In the approximation that this bound applies to pion scattering in real world QCD, we find it gives a window of only ${\sim}150$ MeV for the expected mass of the yet unmeasured spin-7 meson.}

\preprint{LCTP-24-16}

\maketitle

\section{Introduction and Main Results} \label{sec:intro}

The spins and masses of heavy mesons measured in neutral pion scattering can be organized into Regge trajectories, curves along which states with increasing spin have increasing mass. Historically, the study of Regge trajectories lead to the discovery of the Veneziano amplitude \cite{Veneziano:1968yb} and they continue to be essential for understanding many features of scattering amplitudes with massive resonances \cite{Caron-Huot:2016icg,Haring:2023zwu,Eckner:2024ggx}.

Despite their ubiquity in theories with spinning massive states, such as meson scattering or string theory, there is no known formal proof that amplitudes are required to have Regge trajectories. Moreover, it is not known what properties Regge trajectories must satisfy if they do appear. In this paper, we take a step towards understanding the generic features of Regge trajectories using techniques developed in the S-matrix bootstrap program. The S-matrix bootstrap seeks to carve out the space of possible scattering amplitudes using consistency conditions. These assumptions can come in many forms \cite{Dixon:1996wi,Adams:2006sv,Elvang:2013cua,Caron-Huot:2020cmc,Arkani-Hamed:2020blm,Chi:2021mio,Kruczenski:2022lot,Cheung:2023adk}, but, at their core, they rely on the fundamental physical requirements that the S-matrix describes scattering in a unitary, local, causal, and Lorentz-invariant quantum field theory.

In this paper, we employ the so-called ``dual formulation'' of the modern S-matrix bootstrap, which uses dispersion relations to place rigorous exclusion bounds on the space of effective field theories (EFTs) with valid UV-completions \cite{Caron-Huot:2020cmc,Arkani-Hamed:2020blm,Sinha:2020win,Huang:2020nqy,Henriksson:2021ymi,Li:2021lpe,Caron-Huot:2021rmr,Chiang:2021ziz,Chowdhury:2021ynh,Caron-Huot:2022ugt,Albert:2022oes,Fernandez:2022kzi,Albert:2023jtd,CarrilloGonzalez:2023cbf,Berman:2023jys,Albert:2023bml,Haring:2023zwu,Eckner:2024ggx,Berman:2024wyt,Albert:2024yap,Beadle:2024hqg}. With the bootstrap, one can nontrivially test the possible existence of Regge trajectories in the spectrum of the four-point amplitudes in weakly-coupled theories. 

These dispersion relations are used to connect the high and low energy behavior of the amplitude by writing down a four-point amplitude's EFT coefficients in terms of an integral over its high energy spectrum. To ensure the convergence of these dispersion relations, we require that the amplitudes have some minimal integer $n_0$ such that
\begin{align}\label{J0falloff}
    \lim_{|s|\to\infty}\frac{A(s,u)}{s^{n_0}} \to 0
\end{align}
at fixed momentum transfer $u < 0$. This limit, $|s|\to \infty$ with fixed $u < 0$, is called the Regge limit. The Froissart bound suggests that $n_0 \leq 2$ in well-behaved theories satisfying the usual bootstrap assumptions \cite{Froissart:1961ux,Martin:1962rt}. To control the low-energy behavior of the amplitude, we additionally enforce the existence of some maximal spin $J_0$ for the massless states in our theory. 

The analytic version of dual bootstrap emphasized in \cite{Arkani-Hamed:2020blm} and \cite{Chiang:2021ziz} allows us to derive two bounds on the spectra of such amplitudes when the theory is weakly-coupled, crossing symmetric, and the scattered states are massless scalars that are in the adjoint representation of some large-$N$ gauge group. These constraints are:
\begin{enumerate}
\item \textbf{Sequential Spin Constraint} \reef{spinNconst}: States with spin at least as large as $\Jt = \max\{n_0,J_0+1\}$ appear at sequentially increasing masses, i.e. a spin $\ti{J}_1$ state appears at a lighter mass than any spin $\ti{J}_1+1$ state, which is itself exchanged at lower energy than the first $\ti{J}_1+2$ state, and so on. These states form the spectrum's ``leading'' Regge trajectory, the curve along which the lightest state with each spin lies.
\item \textbf{Mass Bound} \reef{massbd}: The lightest state with spin $J \geq \ti{J}_1+2$ which lies on this leading trajectory, has an upper bound determined by the masses and spins of any two states with lower spin on the trajectory.
\end{enumerate}
In the following subsections, we briefly motivate and describe these constraints in more detail and then describe their implications for the physical meson spectrum in the large-$N$ approximation for QCD which agree surprisingly well with experimental measurements.

\subsection{The Sequential Spin Constraint}
A hint that the sequential spin constraint (SSC) should exist was noticed numerically in \cite{Berman:2024wyt}. There, the $n_0 = 0$ case was studied in the $2\to2$ scattering of massless colored scalars with maximal supersymmetry. Simplifying the setup slightly, let us for now consider an $n_0 = 0$ amplitude with crossing symmetry,
\begin{align}\label{crossing}
A(s,u) = A(u,s) \, ,
\end{align}
and has a low-energy expansion of the form
\begin{align}\label{introleexp}
    A(s,u) = \sum_{0\leq q\leq k}^{k_{\max}} a_{k,q}s^{k-q}u^q \quad \quad \quad a_{k,q} = a_{k,k-q} \, .
\end{align}
For the moment, we assume that there are no massless poles. We further require that there be a finite mass gap $M_{\gap}$ below which there are no massive states, a second scale $\mu_{c}M_{\gap}^2$ above the mass gap such that there are no contributions to the imaginary part of the amplitude's spectral density for $M_{\gap}^2 \leq s \leq \mu_c M_{\gap}^2$, and that the amplitude have no $t$-channel poles.\footnote{Were we to take $n_0 = 1$, this would describe scattering of massless pions in large-$N$ QCD \cite{Albert:2022oes,Fernandez:2022kzi,Albert:2023jtd,Albert:2023bml}} Above $\mu_c M_{\gap}^2$, we are agnostic about the spectrum, allowing both single particle exchanges and branch cuts on the real-$s$ axis. The parameter $k_{\max}$ is necessary for numerical implementation and corresponds to truncating the effective Lagrangian at some finite derivative order $2k_{\max}$. Larger truncation order yields stronger bounds.

For the sake of illustration, suppose we take $\mu_c = 1.1$. This choice does not change the results qualitatively for any $\mu_c > 1$. At the mass gap we assume that states with all spins up to some $J_1 > 0$ are exchanged. These exchanges have couplings $g_{J,1}$ for $g_{J,\mu}$ the three-point coupling of two massless scalars to a particle at mass $\mu M_{\gap}^2$ with spin $J$:
\begin{equation}
\label{spinJexch}
\raisebox{-1cm}{\begin{tikzpicture}
	\node at (0.95,1/2) {$g_{J,\mu}^*$};
	\node at (-0.95,1/2) {$g_{J,\mu}^{\phantom{*}}$};
	\node at (0,3/8) {$J$};
	\node at (0,-1/2) {$M^2 = \mu M_{\gap}^2$};
	
	\node (image) at (0,0) {\includegraphics[width=4cm]{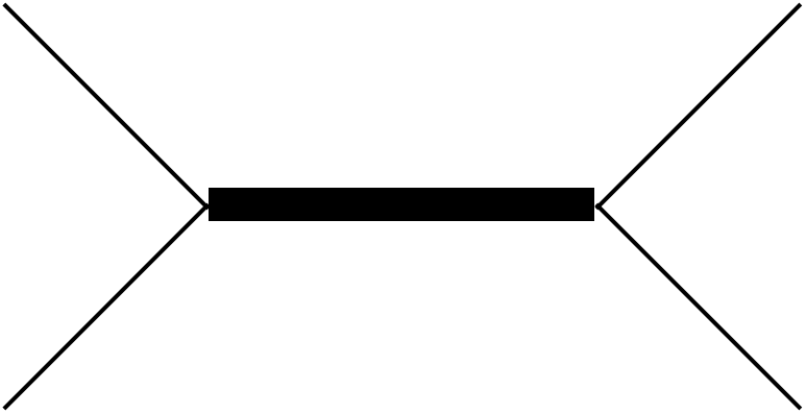}};
\end{tikzpicture}}
\end{equation}
We can numerically determine the maximum allowed value of
\begin{align}
    \frac{|g_{J_1,1}|^2}{a_{0,0}} \,
\end{align}
for $a_{0,0}$ being the coefficient of the four-point contact term by solving a semidefinite optimization problem \cite{Caron-Huot:2020cmc,Albert:2022oes}. To do so, we use SDPB \cite{Simmons-Duffin:2015qma}, an optimization code originally developed for the conformal bootstrap.

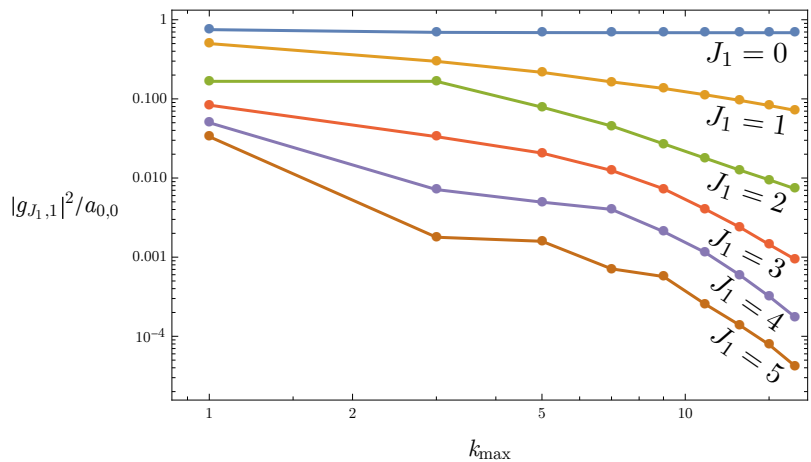
\begin{figure}[t]
\begin{center}
\begin{tikzpicture}
	\node at (4.4,2.4) {$J_1 = 0$};
	\node [rotate = -8] at (4.4,1.5) {$J_1 = 1$};
	\node [rotate = -18] at (4.4,0.55) {$J_1 = 2$};
	\node [rotate = -27] at (4.4,-0.25) {$J_1 = 3$};
	\node [rotate = -35] at (4.4,-0.9) {$J_1 = 4$};
	\node [rotate = -35] at (4.4,-1.6) {$J_1 = 5$};
	
	\node (image) at (0,0) {\includegraphics[width=0.7\textwidth]{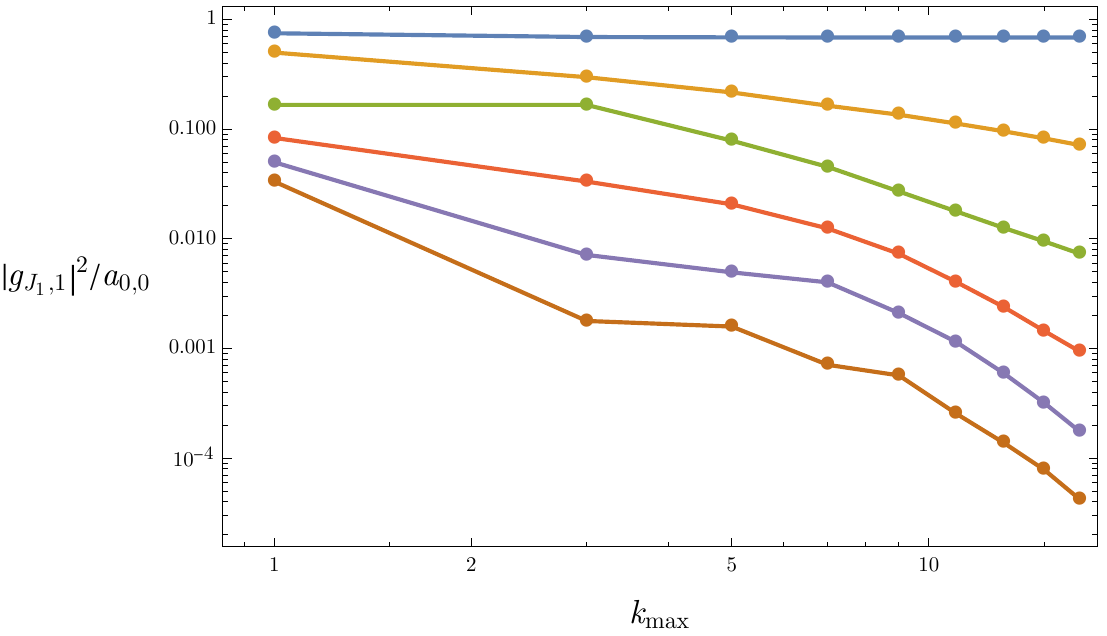}};
\end{tikzpicture}
\end{center}
\caption{A log-log plot of the maximum coupling $|g_{J_1,1}|^2/a_{0,0}$ for states at the mass gap, i.e. the lowest mass state, of spin $J_1 = 0,1,2,3,4,5$ (blue, orange, green, red, purple, brown) to the massless scalars with $\mu_c = 1.1$. While the coupling to the scalar stays constant with increasing $k_{\max}$, the coupling to the state with maximal spin $J_1 > 0$ begins to converge to zero, suggesting that these states are forced to have zero coupling in the large $k_{\max}$ limit so that only scalar exchanges are allowed at the mass gap.}
\label{fig:spincoup0}
\end{figure}

The upper bounds on this ratio for various $J_1$ at increasing truncation orders are shown in Figure \ref{fig:spincoup0}, where we find that the coupling for $J_1 > 0$ converges to zero as we take $k_{\max}$ to be large. The numerical bootstrap appears to be telling us that no states with spin greater than zero are allowed at the mass gap. In \cite{Berman:2024wyt}, a similar analysis showed that at the second mass level, the maximal values for states with spin larger than one have couplings that vanish with increasing $k_{\max}$. We are at most allowed a scalar at the mass gap and a vector at the second mass level, which is precisely the beginning of a leading Regge trajectory.

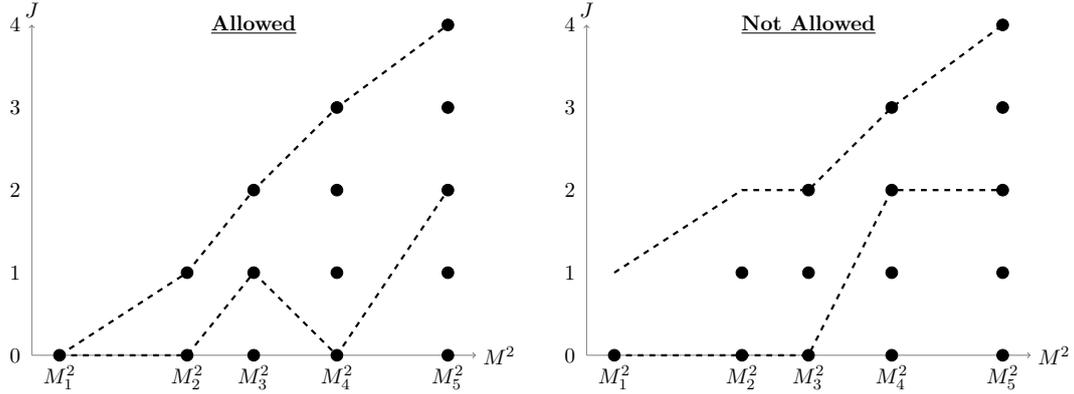
\begin{figure}[t]
	\centering
	\raisebox{2mm}{\scalebox{0.73}{
  \begin{tikzpicture}
  [
    decoration={%
      markings,
      mark=at position 0.5 with {\arrow[line width=1pt]{>}},
    }
  ]
  \draw [help lines,->] (-9,-3) -- (-1,-3) coordinate (xaxis1);
  \draw [help lines,->] (-9,-3) -- (-9,3) coordinate (yaxis1);
  \draw [help lines,->] (1,-3) -- (9,-3) coordinate (xaxis2);
  \draw [help lines,->] (1,-3) -- (1,3) coordinate (yaxis2);

  \node at (-5,3) {\underline{\textbf{Allowed}}};
  \node at (5,3) {\underline{\textbf{Not Allowed}}};

  \node [right] at (xaxis1) {$M^2$};
  \node [above] at (yaxis1) {$J$};
  \node [right] at (xaxis2) {$M^2$};
  \node [above] at (yaxis2) {$J$};

  \node at (-9.3,-3) {$0$};
  \node at (-9.3,-1.5) {$1$};
  \node at (-9.3,0) {$2$};
  \node at (-9.3,1.5) {$3$};
  \node at (-9.3,3) {$4$};

  \node at (0.7,-3) {$0$};
  \node at (0.7,-1.5) {$1$};
  \node at (0.7,0) {$2$};
  \node at (0.7,1.5) {$3$};
  \node at (0.7,3) {$4$};

  \node at (-8.5,-3.4) {$M_1^2$};
  \node at (-6.2,-3.4) {$M_2^2$};
  \node at (-5,-3.4) {$M_3^2$};
  \node at (-3.5,-3.4) {$M_4^2$};
  \node at (-1.5,-3.4) {$M_5^2$};
  
  \node at (1.5,-3.4) {$M_1^2$};
  \node at (3.8,-3.4) {$M_2^2$};
  \node at (5,-3.4) {$M_3^2$};
  \node at (6.5,-3.4) {$M_4^2$};
  \node at (8.5,-3.4) {$M_5^2$};

  \draw [line width=0.4mm,black,dashed] plot coordinates {(-8.5,-3) (-6.2,-1.5) (-5,0) (-3.5,1.5) (-1.5,3)};
  \draw [line width=0.4mm,black,dashed] plot coordinates {(-8.5,-3) (-6.2,-3) (-5,-1.5) (-3.5,-3) (-1.5,0)};
  
  \draw [line width=0.4mm,black,dashed] plot coordinates {(1.5,-1.5) (3.8,0) (5,0) (6.5,1.5) (8.5,3)};
  \draw [line width=0.4mm,black,dashed] plot coordinates {(1.5,-3) (3.8,-3) (5,-3) (6.5,0) (8.5,0)};  

  \draw[fill = black] (-8.5,-3) circle (3pt);
  
  \draw[fill = black] (-6.2,-3) circle (3pt);
  \draw[fill = black] (-6.2,-1.5) circle (3pt);

  \draw[fill = black] (-5,-3) circle (3pt);
  \draw[fill = black] (-5,-1.5) circle (3pt);
  \draw[fill = black] (-5,0) circle (3pt);

  \draw[fill = black] (-3.5,-3) circle (3pt);
  \draw[fill = black] (-3.5,-1.5) circle (3pt);
  \draw[fill = black] (-3.5,0) circle (3pt);
  \draw[fill = black] (-3.5,1.5) circle (3pt);

  \draw[fill = black] (-1.5,-3) circle (3pt);
  \draw[fill = black] (-1.5,-1.5) circle (3pt);
  \draw[fill = black] (-1.5,0) circle (3pt);
  \draw[fill = black] (-1.5,1.5) circle (3pt);
  \draw[fill = black] (-1.5,3) circle (3pt);

  \draw[fill = black] (1.5,-3) circle (3pt);

  \draw[fill = black] (3.8,-3) circle (3pt);
  \draw[fill = black] (3.8,-1.5) circle (3pt);

  \draw[fill = black] (5,-3) circle (3pt);
  \draw[fill = black] (5,-1.5) circle (3pt);
  \draw[fill = black] (5,0) circle (3pt);

  \draw[fill = black] (6.5,-3) circle (3pt);
  \draw[fill = black] (6.5,-1.5) circle (3pt);
  \draw[fill = black] (6.5,0) circle (3pt);
  \draw[fill = black] (6.5,1.5) circle (3pt);

  \draw[fill = black] (8.5,-3) circle (3pt);
  \draw[fill = black] (8.5,-1.5) circle (3pt);
  \draw[fill = black] (8.5,0) circle (3pt);
  \draw[fill = black] (8.5,1.5) circle (3pt);
  \draw[fill = black] (8.5,3) circle (3pt);
  
\end{tikzpicture}
}}

   \caption{\label{fig:RT} A visual representation of the $n_0 = 0$ SSC where $M_N$ is the $N$th mass at which states are exchanged. The points on the first diagonal correspond to the spins saturating the bound $J_N \leq N-1$. The dashed lines on the left correspond to different possible choices of $J_N$ which are allowed by the spin bound. On the right side, they correspond to choices of $J_N$ inconsistent with the spin bound, either because they contain states above the absolute maximum at a given mass level or because they increase in spin too quickly. The leading Regge trajectory can be determined by tracing through the lightest masses at which each spin appears on each dashed line.
   }
\end{figure}

The analytic condition we find exactly matches this numerical intuition that as long as there are only states of finite spin exchanged at the lowest mass level, those states can only be scalars when $n_0 = 0$ and there are no massless poles. We further show a more general constraint which applies at any mass level: for theories with finitely many tree-level exchanges at masses $M_n$ such that $M_1 < M_2 < \cdots < M_N < \cdots$, the largest spin at mass level $n$, which we denote $J_n$,  can only be one higher than the largest spin at any lower mass level. Importantly, this relation only determines the maximum \textit{possible} $J_N$. The spins are trivially at least zero, but have no other lower bound. Therefore, the largest spin at mass level $N$ has to be less than or equal to $N-1$. In Figure \ref{fig:RT}, the dashed lines show various examples of choices for $J_N$ that that are either allowed or disallowed by this bound.

This result can be generalized for any choice of $n_0$ in \reef{J0falloff} and to amplitudes with massless poles, which we ignored in \reef{introleexp}. These massless states affect the dispersive representation of the amplitude that is necessary for argument \cite{Arkani-Hamed:2020blm,Caron-Huot:2022ugt,Albert:2024yap} and can change the strength of the spin bound. The idea that the highest spin allowed increases by at most one at each mass level still applies in the same way, but the starting spin is different. Instead of the maximal spin at the mass gap being zero, it becomes some spin $\Jt$ (the tilde here denotes the maximal \textit{allowed} spin at mass level one, the mass gap, whereas $J_1$ without the tilde is the \textit{actual} maximal spin at the mass gap). The spin $\Jt$ can be determined as 
\begin{align}
\ti{J}_1 = \max\{n_0,J_{0}+1\}
\end{align}
where $J_{0}$ is the maximal spin of the massless states that are exchanged. For example, if the only massless states the external scalars interact with are themselves, $J_0 = 0$. When the massless scalars can also exchange a massless vector, though, the spin-one massless pole makes $J_0 = 1$. Massless graviton exchanges make $J_0 = 2$, but are not directly relevant here because there are no massless spin-two colored particles in the large-$N$ limit, and so with the Froissart bound, this tells us $\Jt \leq 2$. 

With this dependence on Regge behavior and massless poles in mind, we can show that at the lowest mass level $M_{\gap}^2$, we have $J_1 \leq \ti{J}_1$.  As before, the maximal possible spin at higher masses is simply one more than the maximal spin at previous mass levels, so the spin at mass level $N$ always has an absolute maximum of $N+\Jt-1$. We can express this more mathematically as
\begin{align} \label{spinNconst}
    J_{N}\leq \max\{\Jt-1,J_1,J_2,\ldots,J_{N-1}\}+1 \leq N+\Jt-1\, .
\end{align}
In the $n_0 = 0$ case with no massless poles, $\Jt = 0$, which is why the analysis in Figure \ref{fig:spincoup0} shows that only scalars can be exchanged at the mass gap. This bound means that a theory with states that have spin $> \Jt$ must be nontrivially coupled to states of sequentially increasing spin on a leading Regge trajectory.

\subsection{Mass Bound Along the Leading Trajectory}
Consider now a spin $J$ state on the leading Regge trajectory required by the SSC. Crossing symmetry, \reef{crossing}, requires that a general spin $J$ state at mass $\mu M_{\gap}^2$ contribute to the amplitude as
\begin{align}
A(s,u) \supset |g_{J,\mu}|^2\left(\frac{G_J^{(D)}(1+2u/s)}{s-\mu M_{\gap}^2}+\frac{G_J^{(D)}(1+2s/u)}{u-\mu M_{\gap}^2}\right)\,
\end{align}
where $g_{J,\mu}$ is the three point coupling as in \reef{spinJexch} and the $G_J^{(D)}$ are Gegenbauer polynomials. The polynomial $G_J^{(D)}$ is order $J$, so a spin-$J$ state contributes a term of order $s^J$ to the amplitude. Therefore, if a particle with spin $J \geq n_0$ appears in the amplitude, then the amplitude can only diverge slower than $s^{n_0}$ in the Regge limit if the contributions from higher spin states resum such that the Regge behavior is better than the individual terms. Resummation is only possible if we have an infinite tower of massive spinning states so all such theories are \textit{required} to have infinitely many spins exchanged. With the SSC, we can go a step further and find an upper bound on the mass at which states with a particular spin must appear. In its simplest form, with $n_0 = 0$ and no massless poles, the bound tells us that the mass $\wt{M}_J$, where the lightest spin-$J$ state appears, is constrained by $\wt{M}_{1}$, the mass of the lightest spin-one state and $\wt{M}_0 = M_{\gap}$, the mass of the lightest scalar exchanged:
\begin{align}\label{massbd0}
	\left(\frac{\wt{M}_J}{M_{\gap}}\right) \leq \left(\frac{\wt{M}_1}{M_{\gap}}\right)^J\, .
\end{align}
One amplitude that satisfies these assumptions is the open superstring Veneziano amplitude with a particular choice of external scalars\footnote{This amplitude is related by supersymmetry to the original Veneziano amplitude \cite{Berman:2023jys,Berman:2024wyt}.}:
\begin{equation}
\begin{split}
A[zz\bar{z}\bar{z}] =-(\a's)^2A(s,u) \, , \quad \text{where} \quad A(s,u) = \frac{\Gamma(-\a's)\Gamma(-\a'u)}{\Gamma(1+\a't)} \, .
\end{split}
\end{equation}
While the original amplitude $A[zz\bar{z}\bar{z}]$ does not satisfy crossing due to the overall factor of $s^2$, the stripped amplitude $A(s,u)$ does. Further $A(s,u)$ vanishes in the Regge limit ($n_0 = 0$) and, while $A[zz\bar{z}\bar{z}]$ has $J_0 = 1$ due to massless gluon exchanges, the removal of the $s^2$ factor makes $A(s,u)$ act as though it effectively has $J_0 = -1$.\footnote{We are not directly bounding the physical amplitude $A[zz\bar{z}\bar{z}]$ here, so this ``effective'' $J_0$ no longer has the interpretation of being the largest spin of an exchanged massless state. Instead, it is defined by how the massless pole contributes to dispersion relations.} For $A(s,u)$, then, we have
\begin{align}
\Jt^{\text{str}} = \max\{0,-1+1\} = 0\, .
\end{align}
This amplitude saturates the maximal spin constraint with a scalar at the gap, a vector at the second mass level, a spin-two state at the third and so on. The squared-masses are spaced exactly linearly such that $M_2^2 = 2M_{\gap}^2$, $M_3^2 = 3M_{\gap}^2$. The lightest vector appears at $M_{2}^2 = 2M_{\gap}^2$, so $\wt{M}_1 = M_{2} = \sqrt{2}M_{\gap}$. Similarly, $\wt{M}_J = \sqrt{(J+1)}M_{\gap}$. The Veneziano amplitude obeys \reef{massbd0}, since it is always the case that
\begin{align}
\sqrt{J+1} \leq \sqrt{2}^J
\end{align}
for $J \geq 1$.

The bound \reef{massbd0} generalizes to different kinds of Regge behavior and amplitudes with standard massless poles (leading to $\Jt \geq 0$), giving a bound on the mass of the lightest spin $J$ state in terms of the lightest states with two lower spins $J-n_1$ and $J-n_2$. The general expression for the bound is
\begin{align}\label{massbd}
    \frac{\wt{M}_J}{\wt{M}_{J-n_2}} \leq \left(\frac{\wt{M}_{J-n_1}}{\wt{M}_{J-n_2}}\right)^{n_2/(n_2-n_1)} \, \quad \quad \text{for}\quad\quad  n_2 > n_1 > 0 \, ,\, J-n_2 \geq \Jt \,.
\end{align}
The SSC additionally implies the heirarchy $\wt{M}_{J-n_2} < \wt{M}_{J-n_1} < \wt{M}_{J}$. String theory, which characteristically has linear Regge trajectories, satisfies this bound at all spins.

\subsection{Approximate Bounds on QCD}

As a ``real-world'' example, we can consider the spectrum of $\pi\pi\to\pi\pi$ scattering. In the large-$N$ limit of QCD with massless pions, this process satisfies our assumptions with $\Jt = 1$ \cite{Albert:2022oes}. If we ignore any corrections from the fact that real-world QCD has $N = 3$ and that real pions are not massless, we can plug the observed masses of the lowest mass meson resonances, the $\rho$ and $f$ mesons, into \reef{massbd} and find bounds on their masses. This approximation is of course not precise, but our bounds are satisfied for measured values of the mesons. This is quite surprising, especially given that these higher mass resonances are highly unstable, while our bounds only rigorously apply to stable massive states. Further, the authors of \cite{Albert:2023bml} found that, by inputting the measured masses and spins of the $\rho$ and $f_2$ mesons into the numerical bootstrap, there was a corner in the numerical maximum coupling to the $f_2$ meson. This corner suggested the existence of a state with mass ${\sim}1678$\,MeV, extremely close to the physical $\rho_3$ mass! Upon extracting the spectrum of the theory with a state at that mass, they found that the theory had a leading Regge trajectory that had states quite close that of QCD. The compatibility of the large-$N$ QCD bootstrap with real-world QCD suggests that our bounds might still be applicable to real-world QCD even beyond currently measured states, at least in some approximate way.

\begin{table}[t]
\begin{center}
\begin{tabular}{|c c | c c|} 
 \hline
 State & $J$ & Maximum Mass (MeV) & Measured Mass (MeV) \cite{Navas:2024rpp} \\ [0.5ex] 
 \hline
 $\rho(770)$ & 1 & - & $775.3\pm0.2$ \\ [0.5ex] 
 $f_2(1270)$ & 2 & - & $1275\pm1$ \\[0.5ex] 
 $\rho_3(1690)$ &  3 & $m_{f_2}^2/m_{\rho} = 2098$ & $1689\pm 2$ \\[0.5ex] 
 $f_4(2050)$ & 4 & $m_{\rho_3}^2/m_{f_2} = 2237$ & $2018\pm 11$ \\[0.5ex] 
 $\rho_5(2350)$ & 5 & $m_{f_4}^2/m_{\rho_3} = 2411$ & $2330 \pm 35$  \\ [0.5ex] 
 $f_6(2510)$ & 6 & $m_{\rho_5}^2/m_{f_4} = 2690$ & $2470\pm50$ \\ [0.5ex] 
 $\rho_7$ & 7 & $m_{f_6}^2/m_{\rho_5} = 2618$ & ?? \\ [0.5ex] 
 \hline
\end{tabular}
\end{center}
\caption{A table showing the strongest bounds on the experimental masses of different QCD bound states from \reef{massbd} relative to their physical values (the maxima are found using the central experimental values and ignoring error). All measured values lie below the value that would be required by \reef{massbd} if real-world QCD obeyed our formal assumptions. The strongest bounds always come from choosing $n_1 = 1$ and $n_2 = 2$.}
\label{tab:pionbds}
\end{table}

We list the masses and their bounds in Table \ref{tab:pionbds} along with the experimentally determined values for the masses of the states \cite{Navas:2024rpp}. The experimental values are at least 80\% of the maximal values determined by \reef{massbd} and the central experimental value for the $\rho_5$ meson is at 97\% of its maximal value. Assuming the continued validity of the mass bound to real-world QCD continues to hold above the $f_6$ meson, then we can determine an approximate maximal mass for the $\rho_7$ meson using the measured values in Table \ref{tab:pionbds}. The $\rho_7$ meson does not have an experimentally determined mass, but assuming the central values for the masses of the $\rho_5$ and $f_6$ mesons, then the mass bound along with the SSC tell us 
\begin{align}
2470\text{ MeV} = m_{f_6} \leq m_{\rho_7}\leq \frac{m_{f_6}^2}{m_{\rho_5}} = 2618 \text{ MeV} ,
\end{align}
a range of only ${\sim}150$ MeV within which we expect this spin seven state to appear. While the allowed range is quite small, the assumption of large-$N$ implies stable mesons, ignoring the fact that these resonances have decay widths. Given that the $f_6$ meson has a decay width in the tens of MeV, the width of the $\rho_7$ could be similar to this 150 MeV mass window in which we predict that its mass is. It would be interesting if data from collider experiments could be used to further test the applicability of the bounds to QCD for these states with shorter lifetimes.

\subsection{Proof Sketch and Outline}
In Section \ref{sec:aarg}, we prove the SSC for the case shown in Fig. \ref{fig:spincoup0}, with $\Jt = 0$ and no massive loops, then explain how the proof could be straightforwardly generalized. Let us now outline how these above results are derived. 

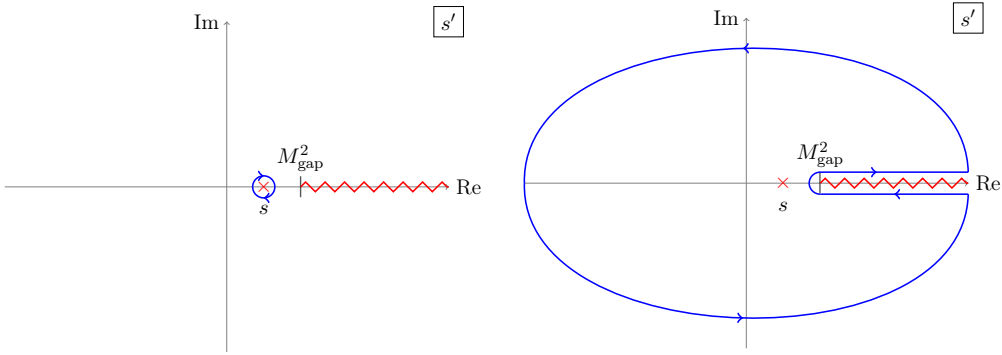
\begin{figure}[t]
	\centering
    \begin{subfigure}[t]{0.45\textwidth}
     \centering
	\raisebox{2mm}{\scalebox{0.73}{
  \begin{tikzpicture}
  [
    decoration={%
      markings,
      mark=at position 0.5 with {\arrow[line width=1pt]{>}},
    }
  ]
  \draw [help lines,->] (-4,0) -- (4,0) coordinate (xaxis);
  \draw [help lines,->] (0,-3) -- (0,3) coordinate (yaxis);
  \node [right] at (xaxis) {Re};
  \node [left] at (yaxis) {Im};

  \node[red] at (2/3,0) {$\times$};
  \node[below=0.2cm] at (2/3,0) {$s$};
  
  \node at (1+1/3,0) {$|$};
  \node[above=0.15cm] at (1+1/3,0) {$M_{\gap}^{2}$};

  \draw[line width=0.8pt, red] (4/3,0) [decorate, decoration=zigzag] --(4,0);
  
  %pole at -t
  \path [draw, line width=0.8pt, postaction=decorate,blue] (2/3-0.2,0) arc (180:0:.2);
  \path [draw, line width=0.8pt, postaction=decorate,blue] (2/3+0.2,0) arc (0:-180:.2);

 \node[draw] at (4,3) {$s'$};

\end{tikzpicture}
}}
    \end{subfigure}\hspace{0mm}\begin{subfigure}[t]{0.45\textwidth}
\centering
\scalebox{0.73}{
\begin{tikzpicture}
  [
    decoration={%
      markings,
      mark=at position 0.5 with {\arrow[line width=1pt]{>}},
    }
  ]
  \draw [help lines,->] (-4,0) -- (4,0) coordinate (xaxis);
  \draw [help lines,->] (0,-3) -- (0,3) coordinate (yaxis);
  \node [right] at (xaxis) {Re};
  \node [left] at (yaxis) {Im};

  \node[red] at (2/3,0) {$\times$};
  \node[below=0.2cm] at (2/3,0) {$s$};
  
  \node at (1+1/3,0) {$|$};
  \node[above=0.15cm] at (1+1/3,0) {$M_{\gap}^{2}$};

  \draw[line width=0.8pt, red] (4/3,0) [decorate, decoration=zigzag] --(4,0);

  \path [draw, line width=0.8pt, postaction=decorate,blue,bend angle=90] (4,0.2) to[bend right] (-4,0);
 \path [draw, line width=0.8pt, postaction=decorate,blue,bend angle=90] (-4,-0) to[bend right] (4,-0.2);
  
  \path [draw, line width=0.8pt, postaction=decorate,blue] (4,-0.2) -- (1+1/3,-0.2);
  \path [draw, line width=0.8pt, postaction=decorate,blue] (1+1/3,-0.2) arc (270:90:0.2) -- (4,0.2);
  
  \node[draw] at (4,3) {$s'$};
\end{tikzpicture}
}
   \end{subfigure} 
   \caption{\label{fig:contour_deformation}
The contour deformation that defines \reef{disprep}. The contour around the branch cut can be identified with the discontinuity of the $s$-channel branch-cut. We do not include them in the in the figure, but there can be an infinite number of massive simple poles on the positive real $s$-axis. These simple poles correspond to exchanges of stable massive states.}
\end{figure}

At the core of the proof are fixed-$u$ dispersion relations. Assuming that our weakly coupled massless $2\to2$ scattering amplitudes $A(s,u)$ are analytic away from the real-$s$ axis, we use the contour deformation shown in Figure \ref{fig:contour_deformation} to relate an amplitude with a mass gap and no $t$-channel cut to the integral over its imaginary part on the positive real $s$ axis.

This contour deformation allows us to write down a dispersion relation for the amplitude at fixed momentum transfer $|u| \ll M_{\gap}^2$ where the amplitude is analytic \cite{Arkani-Hamed:2020blm}:\footnote{The discontinuity of the amplitude is proportional to its imaginary part,  $2i$Im$[A(s,u)] = A(s+i\eps,u)-A(s-i\eps,u)$.}
\begin{align}\label{disprep}
    A(s,u) = A_{\text{sub}} + \int_{M_{\gap}^2}^{\infty} \frac{ds'}{\pi}\frac{\text{Im}[A(s,u)]}{s'-s} \, .
\end{align}
The contribution from the contour at infinity, $A_{\text{sub}}$, vanishes when $n_0 = 0$. Along with the partial wave expansion of a tree-level amplitude,
\begin{align}\label{partwaveexpsketch}
A(s,u) = \sum_{J=0}^{\infty}|g_{J,\mu}|^2G_{J}^{(D)}\left(1+\frac{2u}{s}\right)\, ,
\end{align}
the expression \reef{disprep} and ``$k-q$-times subtracted'' versions of it,
\begin{align}
\frac{A(s,u)}{s^{k-q}} = A_{k-q \text{ sub}} +  \int_{M_{\gap}^2}^{\infty} \frac{ds'}{\pi}\frac{\text{Im}[A(s,u)]}{s^{k-q}(s'-s)} \, ,
\end{align}
allow us to write down dispersion relations for individual Wilson coefficients $a_{k,q}$ by taking contours around $s = 0$. These take the characteristic form
\begin{align}
    a_{k,q} \sim \sum_{J=q}^{\infty}\int_{M_{\gap}^2}^{\infty} ds \frac{\rho_{J}(s)}{s^{k+1}} v_{J,q} \, ,
\end{align}
where $\rho_{J}(s) \geq 0$ encodes information about the massive spectrum. The coefficients $v_{J,q} \geq 0$ are defined in \reef{vlqdef} and come from expanding the Gegenbauer polynomials. In units of the mass gap, tree-level exchanges such as \reef{spinJexch} contribute to Wilson coefficients as
\begin{align}\label{WCtreecont}
    a_{k,q} \supset \frac{|g_{J,\mu}|^2}{\mu^{k+1}}v_{J,q} \, ,
\end{align}
where $M^2 = \mu M_{\gap}^2$ is the mass of the exchanged particle. If we assume that there are only tree-level exchanges at the mass gap with spin up to a finite maximum value $J_{1} > 0$ and some cutoff $\mu_c > 1$  such that $\rho_{J}(s)$ has no support on the interval $(1,\mu_c)$, then the coefficients can be written
\begin{align}\label{sketchakq}
    a_{k,q} \sim \sum_{J = q}^{J_1}|g_{J,1}|^2 v_{J,q} + \sum_{J=q}^{\infty}\int_{\mu_c}^{\infty} ds \frac{\rho_{J}(s)}{s^{k-q+1}} v_{J,q}\, .
\end{align}
We show that in the $k\to \infty$ limit, the integral in \reef{sketchakq} gets suppressed by powers of $\mu_c^{-k}$, so the Wilson coefficients $a_{k,q}$ at large $k$ are simply equal to the sum over contributions at the mass gap. 

We can change integration variables from $s$ to $x = M_{\gap}^2/s$ and find that the dispersive relations for the Wilson coefficients become
\begin{align}
a_{k,q} \sim \sum_{J=q}^{\infty}\int_{0}^{1} dx\, x^k \ti{\rho}_J(x) v_{J,q} \, ,
\end{align}
where $\ti{\rho}_J \propto \rho_J$. As was pointed out in \cite{Arkani-Hamed:2020blm}, this expression means that the $a_{k,q}$ coefficients at fixed $q$ are moments of the distribution $\sum_J m_J(x)v_{J,q}$ and are strongly constrained in terms of each other. One can use these moment bounds along with crossing symmetry \reef{crossing} to find a upper bound on the $q = J_{1}$ Wilson coefficient in terms of a $q = J_{1}+1$ Wilson coefficient for any $J_1 > 0$. In the large $k$ limit, the suppression of the high energy integral, which is the only place there can be dependence on spins larger than $J_1$, means that the $q = J_1$ coefficients depend only on the coupling $|g_{J_1,1}|^2$, while the $q = J_1+1$ coefficient must go to zero as it is independent of the states at the mass gap. Taking the $k\to \infty$ limit implies $|g_{J_{1},1}|^2$ must vanish. There is no upper bound on $a_{k,0}$ in terms of $a_{k,1}$, however, so scalars (and only scalars) can be exchanged at tree-level.

We proceed similarly at general mass level $M_{N}^2$, but with the contributions of tree-level exchanges to the Wilson coefficients given in units of $M_{N}$. Inductively assuming that mass levels below $N$ only contribute spins up to some maximum spin $J-1$, we know Wilson coefficients with $q \geq J$ are independent of any coupling to a state with mass less than $M_{N}$. All of the coefficients with $q \geq J$ then vanish in the large-$k$ limit. We again find an upper bound on $|g_{J_{N},\mu_N}|^2$ (where $\mu_N M_{\gap}^2 = M_N^2$) that goes to zero as $k\to\infty$ unless $J_N \leq J$. The spins must therefore increase sequentially in mass level starting at $J = 0$. Moreover, all spins up to $J$ appear at mass levels below the first level at which a spin-$J$ is exchanged.

Section \ref{sec:NRegTrajs} proves the mass bound \reef{massbd}. The proof is similar to the proof of the SSC, also requiring the moment constraints to show that the coupling to a state which we assume has finite coupling must vanish unless states with higher spin appear below some mass.

In Section \ref{sec:disc}, we give some perspective on our results and suggest possible extensions, while the appendices contain additional proofs necessary for the results in this paper. 

\section{Deriving the Sequential Spin Constraint}\label{sec:aarg}
We begin this section by giving the precise assumptions required for the proof of the sequential spin constraint. We discuss some simple bounds on Wilson coefficients that come from those assumptions and give the rigorous argument for maximal spin at the mass gap. We then explain the generalization of the argument to higher mass levels, but save the technical discussion for Appendix \ref{app:hmasslev}. Finally, we discuss why removing a subset of our assumptions is straightforward and how the spin bound and its proof would be modified without them.

\subsection{Assumptions} \label{sec:assumptions}
We consider planar, color-ordered amplitudes of massless scalars living in the adjoint representation of a large-$N$ group. The behavior of such amplitudes has been bootstrapped, for example, in large-$N$ QCD pion scattering \cite{Guerrieri:2020bto,Albert:2022oes,Albert:2023jtd,Albert:2023bml}, maximally supersymmetric Yang-Mills EFTs \cite{Berman:2023jys,Berman:2024wyt,Albert:2024yap}, and studies of dual-resonant models \cite{Shapiro:1969km,Caron-Huot:2016icg,Cheung:2023adk,Cheung:2023uwn,Saha:2024qpt,Eckner:2024pqt,Cheung:2024uhn}. By considering massless scalar external states and choosing the 1234 color ordering, the amplitudes can be made to be symmetric under $s$ and $u$, recovering the constraint \reef{crossing},
\begin{align}
A[1234]= A(s,u) = A(u,s) \, .
\end{align}
The large-$N$ in the group structure suppresses multi-trace operators such that $A(s,u)$ has no $t$ poles, so we assume that, for fixed $u < 0$, $A(s,u)$ is analytic in the complex $s$ plane anywhere away from the the positive real-$s$ axis. The amplitude must have a partial wave expansion
\begin{align}\label{partwaveexp}
    A(s,u) = \sum_{J = 0}^{\infty}\,n_{J}^{(D)}a_{J}(s)\,G_{J}^{(D)}\Big(1+\frac{2u}{s}\Big) \, ,
\end{align}
in which the sum is over spins $J$ of the exchanged states, $a_{J}(s)$ is the amplitude's spectral density, $n_{J}^{(D)}$ is a dimension-dependent normalization \cite{Correia:2020xtr,Caron-Huot:2020cmc},
and $G_{J}^{(D)}$ are the $D$-dimensional Gegenbauer polynomials,
\be
\label{GegenB}
G^{(D)}_J(x)\equiv {}_{2}{F}_1\left(-J, J + D - 3, \frac{D-2}{2},\frac{1-x}{2}\right)\,.
\ee 
We further require that the massless sector of the theory be weakly coupled so that the contributions to $a_J(s)$ from loops of massless states are suppressed. In this limit, unitarity of the S-matrix reduces to a condition colloquially known as ``positivity'':
\begin{align}\label{posbd}
\text{Im}({a_{J}}(s)) \geq 0 \, .
\end{align}
We assume that, with $n$ finite, at the $n$th mass for which there are tree-level exchanges, there is not an infinite tower of spinning states.
This precludes the existence of amplitudes like $1/(s-m^2)(u-m^2)$, which have non-polynomial residues and have been argued to be nonlocal \cite{Caron-Huot:2016icg,Cheung:2023uwn,Cheung:2024uhn}, though can satisfy all other conditions we require.\footnote{This finite spin constraint does, however, allow amplitudes with accumulation points which have been studied recently \cite{Figueroa:2022onw,Geiser:2022icl,Geiser:2022exp,Maldacena:2022ckr,Cheung:2023adk}.}

Additionally, we enforce the existence of a gap in mass from the massless states to some scale $M_{\gap}$, below which there are no states with nonzero mass, which allows us to write down the fixed-$u$ dispersion relation \reef{disprep} for the amplitude.

The contribution from the contour at infinity in \reef{disprep}, $A_{\sub}$, cannot be accessed in the kind of bootstrap we discuss here, so we typically consider ``$n$-times subtracted'' versions of the dispersion relation \reef{disprep} \cite{Jin:1964zza,Eckner:2024pqt} such that $A_{\text{sub}}$ vanishes:
\begin{align}\label{disprepsub}
    \frac{A(s,u)}{s^{n}} = \int_{M_{\gap}^2}^{\infty} ds'\frac{\text{Im}[A(s,u)]}{s'^{n}(s'-s)} \,.
\end{align}
When the amplitude has the behavior \reef{J0falloff}, any subtraction level $n > n_0$ has no contribution from the contour at infinity. Causality arguments suggest the existence of a Froissart-Martin like bound for polynomially bounded massless amplitudes like \reef{J0falloff} \cite{Camanho:2014apa,Arkani-Hamed:2020blm}
\begin{align}\label{FMbound}
    \lim_{|s|\to\infty}\frac{A(s,u)}{s^2} \to 0 \,
\end{align}
for fixed $-M_{\gap}^2 \ll u < 0$, which guarantees $n_0 \leq 2$.\footnote{It was pointed out in \cite{Albert:2022oes} that because this kind of bound holds for the complex-$s$ plane with fixed $t < 0$ as well as fixed $u$, one can find additional constraints on amplitudes. This property does not play a role in our argument.}
Such a bound is rigorous for the scattering of massive theories \cite{Froissart:1961ux,Martin:1962rt}, but is unproven for massless scattering. 

To simplify the argument necessary for Section \ref{sec:atmgap}, we further constrain the problem. First, we assume that the entire theory, including the massive sector, is weakly coupled such that we are studying the tree-level approximation of our amplitude. This approximation applies naturally, for example, to large-$N$ QCD \cite{Witten:1979kh,Albert:2022oes}. Having the entire theory be weakly coupled suppresses loops of massless states and decays of massive states and ensures that after each massive pole at $s = \mu M_{\gap}^2$ (for $\mu \geq 1$) at which states are exchanged, the amplitude is analytic in $s$ until $s = (\mu+\d) M_{\gap}^2$ for $\d > 0$. 

Second, instead of the standard Froissart bound \reef{FMbound}, we impose softer Regge behavior than is strictly necessary,
\begin{align}\label{impReg}
\lim_{|s|\to\infty}A(s,u) \to 0 \,
\end{align}
for fixed $u < 0$, taking $n_0= 0$ in \reef{J0falloff}. 

Third, we require that there are no massless poles in the low-energy expansion of the amplitude:
\begin{align} \label{LEexp}
A(s,u) = \sum_{0\leq q\leq k}^{\infty} a_{k,q} s^{k-q}u^{q} \quad \quad \quad a_{k,q} = a_{k,k-q}\, .
\end{align}
This is not necessarily the case for any particular theory, but, along with \reef{impReg}, simplifies our discussion. The precise argument given here applies, with a small change in notation, to massless scalar scattering in $\Nd = 4$ supersymmetric theories \cite{Berman:2023jys,Berman:2024wyt}. All three of these conditions can be weakened and the argument would still apply with only slight modification, as we describe in Sec. \ref{sec:remsimp}.

\subsection{Dispersion Relations and Simple Consequences}
We can use the contour deformation in Figure \ref{fig:contour_deformation} and the partial wave expansion \reef{partwaveexp} to rewrite the $(k-q)$-times subtracted dispersion relation as
\begin{align}
    \oint\frac{ds'}{2\pi i}\frac{A(s,u)}{s'^{k-q}(s'-s)} = \sum_{J = 0}^{\infty}\int_{M_{\gap}^2}^{\infty} \frac{ds'}{\pi} \frac{n_{J}^{(D)}\text{Im}[a_{J}(s')]}{s'^{k-q}(s'-s)}G_{J}^{(D)}\Big(1+\frac{2u}{s}\Big) \, 
\end{align}
for $D > 3$. Then, taking $s \to 0$, using the low-energy expansion \reef{LEexp}, applying $q$ derivatives with respect to $u$, and setting $u \to 0$, we find
\begin{align}\label{disprepinit}
    a_{k,q} = \frac{1}{q!}\frac{\partial^q}{\partial u^q}\left(\sum_{J = 0}^{\infty}\int_{M_{\gap}^2}^{\infty} \frac{ds}{\pi} \frac{n_{J}^{(D)}\text{Im}[a_{J}(s)]}{s^{k-q+1}}G_{J}^{(D)}\Big(1+\frac{2u}{s}\Big)\right)\Bigg|_{u = 0} \,.
\end{align}
The only $u$-dependence in the expression comes from the Gegenbauer polynomaials, so, defining the coefficients $v_{\ell,q}^{(D)}$ such that
\begin{align}\label{vlqdef}
  G_J^{(D)}(1+2\delta)
    &=
    \sum_{q=0}^J
    v_{J,q}^{(D)} \, 
    \delta^n
    \, , & v_{J,q}^{(D)}
    &=
    \binom{J}{q}
    \frac{ \Gamma(J+D-3+q) \, \Gamma(\frac{D-2}{2}) }
         { \Gamma(J+D-3) \, \Gamma(\frac{D-2}{2}+q) }
    \, ,
\end{align}
we find
\begin{align}
    a_{k,q} = \sum_{J = 0}^{\infty}\int_{M_{\gap}^2}^{\infty} \frac{ds}{\pi} \frac{n_{J}^{(D)}\text{Im}[a_{J}(s)]}{s^{k+1}} v_{J,q}^{(D)} \,.
\end{align}
From here on we drop the superscript $(D)$ as our argument is dimension independent. The improved Regge behavior \reef{impReg} allows us to write dispersion relations for all $a_{k,q}$ Wilson coefficients of \reef{LEexp}. Upon rescaling\footnote{We can equivalently think of this as choosing units in which $M_{\gap}^2 = 1$.}
\begin{align}\label{rescale0}
    (M_{\gap}^2)^{k+(D-4)/2} a_{k,q} \to a_{k,q}
\end{align}
to make the Wilson coefficients dimensionless, we find \cite{Berman:2023jys,Berman:2024wyt}:
\begin{equation}
\label{finalequ}
a_{k,q}\,=\,\sum_{J=0}^{\infty}\int_{1}^\infty dy \,f_{J}(y) \,y^{-k} v_{J,q}
\end{equation}
where $y = s/M_{\gap}^2$ is the dimensionless center-of-mass energy and the normalized spectral density is $f_J(y) = y^{-(D/2+1)} n_J \text{Im}(a_{J}(M_{\gap}^2y))/\pi \ge 0$. 

The two properties of $v_{J,q}$ important here are that 
\begin{equation}
\begin{split}
&v_{J,q} \geq 0 \text{ for all $J$ and $q$}\\
&v_{J,q} = 0 \text{ for } J < q \, .
\end{split}
\end{equation}
This second property means that we can make the lower bound in the sum over $J$ depend on $q$,
\begin{align} \label{disprepq}
a_{k,q} = \sum_{J = q}^{\infty}\int_{1}^\infty dy \,f_{J}(y) \,y^{-k} v_{J,q} \, ,
\end{align}
so, naively, $a_{k,q}$ does not depend on the exchange of states with spin $J < q$. For notational convenience we define the bracket
\be 
 \label{defbracket}
\Big\la y^{-k}v_{J,q}\Big\ra_\mu
= \sum_{J=q}^{\infty}\int_{\mu}^\infty dy \,f_{J}(y) \,y^{-k}v_{J,q} \, .
\ee
Then
\begin{align}
    a_{k,q} = \left\la y^{-k} v_{J,q} \right\ra_1 \, .
\end{align}
The positivity of the integrand in \reef{disprepq} and the fact that $y \geq 1$ trivially tells us that 
\begin{align}\label{basicbd}
a_{k,q} \geq a_{k',q} \quad \text{for} \quad k \leq k' \, .
\end{align}
Along with the crossing constraint \reef{crossing}, which requires
\begin{align}\label{akqcrossing}
    a_{k,q} = a_{k,k-q}
\end{align}
for all coefficients, we can show that \cite{Berman:2023jys}
\begin{align}\label{barbd}
    0 \leq \bar{a}_{k,q} \equiv \frac{a_{k,q}}{a_{0,0}} \leq 1 \, ,
\end{align}
for all $k,q$. In \cite{Arkani-Hamed:2020blm}, finding bounds on the $a_{k,q}$ coefficients was identified with a moment problem and, among several other conditions, were found to obey a so-called ``Hankel matrix'' constraint, with
\begin{align}\label{hankmat}
\begin{pmatrix}
a_{q,q} & a_{q+1,q} & \ldots\\
a_{q+1,q} & a_{q+2,q} & \ldots \\
\vdots & \vdots & \ddots
\end{pmatrix} \, 
\end{align}
being totally nonnegative, meaning each minor must have nonnegative determinant. In Appendix \ref{app:Hankbd}, we show a particular consequence of \reef{hankmat} is that
\begin{align}\label{hankbd}
\left(\frac{a_{k,q}}{a_{q,q}}\right)^{(k'-q)/(k-q)} \leq \left(\frac{a_{k',q}}{a_{q,q}}\right) \, 
\end{align}
for $k \leq k'$.

A tree-level exchange at mass $\mu M_{\gap}^2$ of a state with spin $J$ contributes to the imaginary part of the spectral density as
\begin{align}\label{spectocoup}
    n_{J}\text{Im}(a_{J}(s))
    \,\supset\,
\frac{|g_{J,\mu}|^2}{\big(\mu M_{\gap}^2\big)^{(4-D)/2}} \d\big(s/M_{gap}^2-\mu\big) \,,
\end{align}
where $g_{J,\mu}$ is the (appropriately normalized) dimensionless coupling of the massless states to to the massive state. With the rescaling \reef{rescale0}, the contribution of a tree-level exchange to a Wilson coefficient is
\begin{align}\label{treeakqcont}
    a_{k,q} \supset \frac{|g_{J,\mu}|^2}{\mu^{k+1}} \, .
\end{align}
We now have all the necessary ingredients to prove for this setup that only scalars can be exchanged at the mass gap.

\subsection{At the Mass Gap}\label{sec:atmgap}
There are two main steps in the proof. First, we show that, in the large $k$ limit, any fixed-$q$ Wilson coefficient $a_{k,q}$ is completely determined by contributions at the mass gap. This means that upon assuming the existence of some maximal spin exchanged at $M_{\gap}^2$, $J_{1}$, any Wilson coefficient with $q \geq J_{1}$ must vanish in the $k\to\infty$ limit since coefficients with $q = J$ have only contributions from states with spin $\geq J$. Further, the large-$k$ limit of the coefficient with $q = J_{1}$, $a_{k,J_{1}}$, is exactly proportional to the coupling to the state with spin $J_{1}$ at the mass gap. Second, we use the Hankel matrix constraints to find an upper bound on $a_{k,J_{1}}$ in terms of a coefficient with $q,k-q > J_{1}$ that vanishes in the large-$k$ limit, implying that the coupling to the state with spin $J_{1}$ at the mass gap must also be zero.

The first step relies on the weak coupling limit of the massive sector. In this approximation, there are no contributions to the spectrum with mass-squared between $M_{\gap}^2$ and some cutoff scale $\mu_c M_\gap^2$ for $\mu_c > 1$ where the next tree-level exchanges appear. The spectrum $f_J(y)$ has no support for $y\in(1,\mu_c)$ and we can write the dispersion relation \reef{disprepq} as
\begin{align} \label{disprepqMG}
a_{k,q} = \sum_{J = q}^{J_{1}}|g_{J,1}|^2v_{J,q} + \la y^{-k}v_{J,q}\ra_{\mu_c} \, .
\end{align}
Assuming $k \geq 2J_{1}+1$ so that $J_{1} < k-J_{1}$, we use \reef{disprepqMG} to find from crossing 
\begin{align}\label{crossdisp0}
a_{k,k-J_{1}} = a_{k,J_{1}} = |g_{J_{1},1}|^2v_{J_{1},J_{1}} + \la y^{-k}v_{J,J_{1}}\ra_{\mu_c} \, . 
\end{align}
Dividing by $a_{0,0}$, we get
\begin{align}\label{crossdisp0bar}
\bar{a}_{k,k-J_{1}} = \bar{a}_{k,J_{1}}= |\bar{g}_{J_{1},1}|^2v_{J_{1},J_{1}} + \frac{1}{a_{0,0}}\la y^{-k}v_{J,J_{1}}\ra_{\mu_c} \, ,
\end{align}
where $|\bar{g}_{J,M^2}|^2 \equiv |g_{J,M^2}|^2/a_{0,0}$. All quantities in this expression must be finite due to \reef{barbd}. Consider the integral over contributions above $\mu_c$: for $y\in[\mu_c,\infty)$ and $k \geq J_1$, we know
\begin{align}
y^{-k} = \mu_c^{-k}\left(\frac{y}{\mu_c}\right)^{-k} \leq \mu_c^{-k}\left(\frac{y}{\mu_c}\right)^{-J_{1}},
\end{align}
so, at any $J$, the integrand obeys
\begin{align}
y^{-k}v_{J,q}f_{J}(y) \leq \frac{y^{-J_{1}}}{\mu_c^{k-J_{1}}}v_{J,q}f_{J}(y) \, .
\end{align}
Since we assumed $k\geq 2J_{1}+1$,  we then get the inequality
\begin{align}\label{HEintbd}
\frac{1}{a_{0,0}}\la y^{-k}v_{J,J_{1}}\ra_{\mu_c}  \leq \frac{1}{\mu_c^{k-J_{1}}}\frac{1}{a_{0,0}}\la y^{-J_{1}}v_{J,J_{1}}\ra_{\mu_c}\, .
\end{align}
Using \reef{disprepqMG}, we can see the high energy integral on the right hand side of \reef{HEintbd} is equal to $\bar{a}_{J_{1},J_{1}}-|\bar{g}_{J_{1},1}|^2$, so we find
\begin{align}\label{hevan1}
\frac{1}{a_{0,0}}\la y^{-k}v_{J,J_{1}}\ra_{\mu_c}  \leq \frac{\bar{a}_{J_{1},J_{1}}-|\bar{g}_{J_{1},1}|^2}{\mu_c^{k-J_{1}}} \, .
\end{align}
Since $\bar{a}_{J_{1},J_{1}}-|\bar{g}_{J_{1},1}|^2 \leq 1$ and $\mu_c > 1$, as we take $k\to\infty$ the right hand side of \reef{hevan1} goes to zero, meaning the high-energy integral on the left hand side of \reef{hevan1} must vanish as well. Taking the $k\to\infty$ limit of \reef{crossdisp0bar}, we then see
\begin{align}\label{singcouplim}
\lim_{k\to\infty}\bar{a}_{k,J_{1}} = |\bar{g}_{J_{1},1}|^2v_{J_{1},J_{1}} \, .
\end{align}
We have found, as desired, a Wilson coefficient which depends entirely on the coupling to the spin $J_{1}$ state at the mass gap. We can use this same idea to show that there exist Wilson coefficients which vanish in the large $k$ limit. In particular, since $k > 2J_{1}+1$, the sum over spins in \reef{disprepq} starting at $J = q$ implies neither $\bar{a}_{k+1,k-J_{1}}$ nor its crossing symmetric partner, $\bar{a}_{k+1,J_{1}+1}$, have contributions from the states in the tower at the lowest mass level. Therefore, we can bound $\bar{a}_{l+1,k-J_1}$ from above by using crossing symmetry,
\begin{equation}\label{kmjmaxvanish}
\begin{split}
\bar{a}_{k+1,k-J_{1}} = \bar{a}_{k+1,J_{1}+1}&= \frac{1}{a_{0,0}}\la y^{-(k+1)}v_{J,J_{1}+1}\ra_{\mu_c} \\
&\leq \frac{\bar{a}_{J_{1}+1,J_{1}+1}}{\mu_c^{k-J_{1}}} \,.
\end{split}
\end{equation}
Since $\bar{a}_{J_{1}+1,J_{1}+1}$ is finite, it must be the case that $\bar{a}_{k+1,J_{1}+1}$ vanishes in the large-$k$ limit. 

We can now move on to the second step necessary to prove only scalar exchanges are allowed at the mass gap: showing that the coupling to the spin $J_{1}$ state must be zero if $J_1>0$ by finding a bound on $\bar{a}_{k,k-J_{1}}$ in terms of $\bar{a}_{k+1,k-J_{1}}$ and taking the $k\to\infty$ limit. We can get this bound by setting $q = k - J_{1}$, $k' = k+1$ in \reef{hankbd}:
\begin{align}\label{hankbdvan}
\left(\frac{a_{k,k-J_{1}}}{a_{k-J_{1},k-J_{1}}}\right)^{1+1/J_{1}} \leq \left(\frac{a_{k+1,k-J_{1}}}{a_{k-J_{1},k-J_{1}}}\right) \, .
\end{align}
Barring all coefficients and using crossing, we can write $\bar{a}_{k+1,k-J_{1}} = \bar{a}_{k+1,J_{1}+1}$ and $\bar{a}_{k-J_{1},k-J_{1}} = \bar{a}_{k-J_{1},0}$, which has the dispersive representation
\begin{align}\label{spechank1}
\bar{a}_{k-J_{1},0} = \sum_{J = 0}^{J_{1}}|\bar{g}_{J,1}|^2v_{J,0} + \frac{1}{a_{0,0}}\la y^{J_{1}-k}\ra_{\mu_c}.
\end{align}
As we saw in \reef{HEintbd}, the high energy integral is bounded by $1/\mu_c^{k}\times$finite, so this coefficient is equal to just the sum over the couplings in the large $k$ limit. Taking the large $k$ limit in the barred version of \reef{hankbdvan}, we obtain
\begin{align}\label{hankbdvanlim}
\left(\frac{|\bar{g}_{J_{1},1}|^2v_{J_{1},J_{1}}}{\sum_{J = 0}^{J_{1}}|\bar{g}_{J,1}|^2 }\right)^{1+1/J_{1}} \leq \frac{\lim_{k\to\infty}\bar{a}_{k+1,J_{1}+1}}{\sum_{J = 0}^{J_{1}}|\bar{g}_{J,1}|^2} = 0\, ,
\end{align}
where we have used the fact that $v_{J,0} = 1$ for all $J$. Then, since $v_{J_{1},J_{1}}$ is finite for finite $J_{1}$, \reef{hankbdvanlim} implies
\begin{align}
\frac{|\bar{g}_{J_{1},1}|^2}{\sum_{J = 0}^{J_{1}}|\bar{g}_{J,1}|^2 } = 0\, .
\end{align}
Thus, the coupling to spin $J_{1}$ has zero contribution, so the theory can have a tower of states only up to $J_{1} - 1$. This argument holds for any integer $J_1 \geq 1$, so rules out exchanges of any states with spin larger than or equal to one. At $J_{1} = 0$, though, it fails, perhaps most obviously because the Hankel constraint \reef{hankbd} can be satisfied with $a_{q+n,q}/a_{q,q} = 0$ for all coefficients with integer $n > 0$, so it has no bounding power on the scalar coupling. Therefore, we have shown that $J_1 = 0$, i.e.~that at the lowest mass level of any tree-level UV completion of a colored scalar theory which admits zero-times subtracted dispersion relations, there can only be a massive scalar exchanged.

\subsection{Higher Mass Levels}\label{sec:hmass}

We can modify the proof in the previous section to show that the largest spin of a state that could possibly be exchanged at any mass level above the mass gap is one higher than the highest spin exchanged at lower mass levels. That is to say, at the second mass level, we can exchange at most a spin-one state since we can have only spin-zero states at the gap, at the third mass level, we can exchange at most a spin-two state if a vector is exchanged at the second mass level or at most a vector if only scalars are exchanged at the second mass level, and so on. Using $J_N$ to denote the largest spin exchanged at mass level $n$, this constraint tells us that the maximal spin exchanged at mass level $N$ obeys
\begin{align}\label{maxJn}
J_N \leq \ti{J}_{N} \, .
\end{align}
where we are using the shorthand\footnote{We know $J_1 = 0$ because we chose the masses $M_N$ such that there are states exchanged at $s = M_{N}^2$ and Section \ref{sec:atmgap} proved that $J_1 \leq 0$}
\begin{align}
\ti{J}_{n} = \max\{-1,J_1 = 0,J_2,\ldots,J_{n-1}\} + 1 \,
\end{align}
for spin one larger than the maximal spin exchanged at any level below $N$. Note that we do not require $J_N = \ti{J}_N$, so we can think of $\ti{J}_N$ as the maximum \textit{possible} spin exchanged at mass level $N$, while $J_N$ is the \textit{actual} maximal spin exchanged at mass level $N$. To find this maximal spin constraint, we use the same strategy we employed in the previous section: first find the coefficient which isolates the contribution of a the maximal spin-$J_N$ state at some mass level $N$ and then use the Hankel matrix bounds to show that, if $J_N > \ti{J}_{N}$, the coupling to the spin-$J_N$ state must vanish. 

In this language, Section \ref{sec:atmgap} proves that $J_1 = \ti{J}_1 = 0$. At any mass level below the second (i.e.~only at the first mass level), then, the exchanged states are maximally spin zero. This is the base case for the inductive assumption that below mass level $N$, only states with spin up to a maximum $\ti{J}_{N}-1$ can be exchanged. We now want to show that, if this is true up to some mass level $N-1$, then $J_{N} \leq \ti{J}_{N}$.

We reserve this proof for Appendix \ref{app:hmasslev} because the core idea is identical to that in Section \ref{sec:atmgap} and just requires more detailed notation. The only additional insight is that if we consider rescaled coefficients
\begin{align}\label{rescaleakq}
\mu_N^{k+1} a_{k,q}\,,
\end{align}
then we compensate for the mass suppression of the states at mass level $N$ in the $k\to \infty$ limit, and so those contributions from exchanges with $s = \mu_NM_{\gap}^2$ become finite. While this also makes contributions from states at levels $n<N$ infinite in the large $k$ limit, our inductive assumption tells us that those states all have spins only up to $\ti{J}_{N}-1$, and so do not contribute to rescaled Wilson coefficients $\mu_Na_{k,q}$ with $q,k-q \geq \ti{J}_{N}$ due to the properties of the $v_{J,q}$ constants. This identifies finite rescaled coefficients which have contributions only from the state with spin $J_N$, which we can then bound from above by zero using the Hankel matrix constraints if $J_N > \ti{J}_{N}$. 

These bounds result in us inductively showing that crossing symmetry and the Regge behavior $\lim_{|s|\to\infty}A(s,u) \to 0$ imply that any amplitude subject to our assumptions has a maximal spin of $\ti{J}_{N}$ at mass level $N$ in its spectrum. As a consequence, the highest spin in the massive spectrum must increase one at a time: an amplitude having exchanges of states with spin $J$ at some mass necessitates the existence of states with spin $J-1$ at some lower mass, spin $J-2$ at an even lower mass, and so on until we reach the scalar at the mass gap. Together, these conditions tell us the maximal allowed spin at mass level $N$ has the bound
\begin{align}\label{pvSSC}
    J_N \leq \ti{J}_{N} = \max\{-1,J_1,J_2,\ldots,J_{N-1}\} +1 \leq N-1\, ,
\end{align}
exactly matching the constraint \reef{spinNconst} with $\Jt = 0$.

\subsection{Removing Simplifying Assumptions}\label{sec:remsimp}

In addition to the assumptions which were strictly necessary for the proof of the sequential spin constraint to work, we made three simplifications to the problem in Section \ref{sec:assumptions}:
\begin{enumerate}
\item The entire theory, both massless and massive sectors, is weakly coupled so that the tree-level approximation is valid at all energy scales.
\item The amplitude vanishes in the Regge limit, i.e. that $n_0 = 0$ in \reef{J0falloff}.
\item The low-energy expansion of the amplitude has no massless poles.
\end{enumerate}
In this section, we discuss to what extent these assumptions can be relaxed and how the proof would need to be modified.

Loosening the weak coupling assumption in the massive sector (though keeping it for the massless sector) to allow massive loops means there are no longer gaps in the support of $\text{Im}(a_{J}(s))$ above some scale $\mu_{\lop}M_{\gap}^2$ at which the loops first appear. Further, loop level contributions are not purely polynomial in Mandelstams, so cannot be approximated by a finite sum over Gegenbauer polynomials. They contribute to the spectral density at all $J$, no matter the spin of the state in the loop, violating the assumption that we can pick some maximal spin $J_N$ for some mass level $N$ above which $\text{Im}(a_{J}(s)) = 0$ \cite{Arkani-Hamed:2020blm}. However, so long as the loop-level contributions to the Wilson coefficients are finite, then they would be suppressed at large $k$ because their contributions inherently come from integrating over the real $s$-axis above some threshold and large $s$ information comes with factors of $(s/M_{\gap}^2)^{-k}$. Thus, if $\mu_{\text{loop}} > 1$, then the argument holds for all tree-level exchanges below $\mu_{\text{loop}}M_{\gap}^2$. 

When the massless sector is not weakly coupled, then the amplitude can have loops of massless states, leading to branch cuts that reach down to $s = 0$ in the complex $s$ plane. We are then unable to define our naive Wilson coefficients \cite{Arkani-Hamed:2020blm}. As long as the theory remains perturbative, the massless loop contributions can be determined in terms of the tree-level Wilson coefficients, so we could expect that these bounds might at least be approximately true, but they would not rigorously apply.

Relaxing the Regge behavior constraint to general $n_0$ in \reef{J0falloff} means that we need to take at least $n_0$ subtractions for $A_{\text{sub}}$ to vanish in \reef{disprepsub}, and so we have convergent dispersion relations only for $a_{k,q}$ with $q,k-q\geq n_0$. This is exactly analogous to how we consider only rescaled coefficients with $q,k-q \geq \ti{J}_{N}$ in Section \ref{sec:hmass}. The base case for the inductive argument at the first mass level follows the same logic as if we took $N = n_0$ in Section \ref{sec:hmass}, and so we rule out any massive states with spins $J > n_0$ being exchanged at $M_{\gap}^2$.\footnote{With the Froissart bound \reef{FMbound}, we additionally know $n_0\leq 2$.} This, for example, explains why the massive spin-one $\rho$ meson is found at the lowest mass level in pion scattering \cite{Navas:2024rpp} and a trajectory with $J = n$ is found for the at least singly-subtracted dispersion relations pertinent to large-$N$ QCD \cite{Albert:2022oes,Albert:2023bml}.

A similar phenomenon occurs if we allow for the existence of massless poles. The low energy expansion, \reef{LEexp}, then becomes
\begin{align}
A(s,u) = g^2\left(\frac{f(u)}{s}+\frac{f(s)}{u}\right) + \sum_{0\leq q \leq k}^{\infty}a_{k,q}s^{k-q}u^q\, .
\end{align}
If the only massless states the scalars interact with are themselves, then $f(s) = 1$ and the zero-times subtracted dispersion relation picks up an additional contribution from the $1/u$ pole which goes to infinity in the $u \to 0$ limit we use in the dispersive representation of the Wilson coefficients \reef{disprepinit}. These massless contributions interfere with the crossing relations and Hankel bounds.\footnote{See Sec. 4 in \cite{Arkani-Hamed:2020blm}.} The fact that they are not polynomial in $u$ means they require ``smeared'' dispersion relations \cite{Caron-Huot:2021rmr,Albert:2024yap,Beadle:2024hqg}. These smeared relations do not allow the same kind of Hankel type bounds essential for our argument, so we would need to start with once-subtracted dispersion relations even if $n_0 = 0$. The beta function amplitude,
\begin{align}
    A(s,u) = \frac{\Gamma(-s)\Gamma(-u)}{\Gamma(-s-u)},
\end{align}
relevant in bootstraps for amplitudes with dual resonance \cite{Cheung:2023adk,Cheung:2023uwn,Cheung:2024uhn}, has this property. Despite having the vanishing Regge behavior of \reef{impReg}, it has a spin-one state at its first mass level. This is because of its scalar massless pole, which means that we cannot write dispersive representations for the Wilson coefficients $a_{k,k}$ on their own.

More generally, if the massless scalars interact with massless states of spin $J_0$, then $f(s) \supset s^{J_0}/u$, so we need dispersion relations which are subtracted at least $J_0 +1$ times to get rid of their contributions. Together, then, the falloff condition and massless pole conditions combine to necessitate at least
\begin{align}
    \Jt = \max\{n_0,J_0\}\,
\end{align}
subtractions, and so we replace the $-1$ in \reef{pvSSC} with $\Jt-1$ to get 
\begin{align}\label{pvSSCfull}
    J_N \leq \max\{\Jt-1,J_1,J_2,\ldots,J_{N-1}\} +1 \leq N+\Jt-1\,,
\end{align}
where the second inequality comes from the fact that the spins can increase at most by one at each level starting at spin-$\Jt$ at mass level one. This is precisely \reef{spinNconst}, the sequential spin constraint. 

\section{Finding Upper Bounds on Masses} \label{sec:NRegTrajs}
To show the power of the sequential spin constraint when combined with the moment-type constraints, we prove in this section that, together, they place an upper bound on the lightest mass at which there are states of a given spin in terms of the lightest masses at which there are lower spins. 

Let $\ti{\mu}_J$ be defined such that it is the first mass at which a state of spin $J$ appears. The SSC tells us that there is a sequence $\ti{\mu}_{\ti{J}_1} < \ti{\mu}_{\ti{J}_1+1} < \ldots < \ti{\mu}_J < \ldots < \ti{\mu}_{J_{E}} < \mu_{\lop}$ for some finite masses $\mu_{J}$ below $\mu_{\lop}$ and $J_{E}$ the highest spin for a particle exchanged below the loop scale.\footnote{These masses $\ti{\mu}$ are the exchanges on the theory's leading Regge trajectory.} In this section, we directly show that these masses obey the bound
\begin{align}\label{massbdJ}
    \frac{\ti{\mu}_{J+n}}{\ti{\mu}_{\Jt}} \leq \left(\frac{\ti{\mu}_J}{\ti{\mu}_{\Jt}}\right)^{(J+n-\Jt)/(J-\Jt)} \,.
\end{align}
for $J \geq \Jt$ and $0 \leq n \leq J_E - J$. It is possible to prove a generalization of \reef{massbdJ} simply by replacing $\ti{J}_1$ with $\ti{J}$ for any $\ti{J} < J$ in the argument we present.\footnote{We thank Nick Geiser for pointing this out.} 
The most general bound then reads
\begin{align}\label{massbdJgen}
    \frac{\ti{\mu}_{J+n}}{\ti{\mu}_{\ti{J}}} \leq \left(\frac{\ti{\mu}_J}{\ti{\mu}_{\ti{J}}}\right)^{(J+n-\ti{J})/(J-\ti{J})} \, .
\end{align}
Making the replacements $\ti{J} \to J-n_2$, $J+n\to J$, and $J \to J-n_1$ in this expression, remembering that $\ti{\mu}_N$ is defined such that $\wt{M}_N^2 = \mu_N M_{\gap}^2$, and taking the square root of both sides of this expression, we find
\begin{align}\label{massbdJgen1}
    \frac{\wt{M}_{J}}{\wt{M}_{J-n_2}} \leq \left(\frac{\wt{M}_{J-n_1}}{\wt{M}_{J-n_2}}\right)^{n_2/(n_2-n_1)} \, .
\end{align}
for $0\leq n_1 < n_2$, reproducing the constraint given in \reef{massbd}. 

The argument for this bound is technical for unfixed $\Jt$, but displays some of the relevant complications with the more general proof. The core idea that we use the combination of crossing symmetry and the Hankel matrix bounds remains the same, with the main requirement being more difficult notation.

To start, we use \reef{treeakqcont} to write the Wilson coefficients for amplitudes with tree-level exchanges up to $\mu_{\lop}$ as 
\begin{align}\label{treelevdisp}
    a_{k,q} = \sum_{n = 1}^{N}\sum_{J = q}^{\infty}\frac{|g_{J,\mu_n}|^2v_{J,q}}{\mu_n^{k+1}}+\la y^{-k}v_{J,q}\ra_{\mu_{\lop}}\, ,
\end{align}
where $\mu_N < \mu_{\lop}$ is the largest mass level at which states are exchanged below the scale at which loops contribute to the mass spectrum of the theory $\mu_{\lop}M_{\gap}^2$. 
Now we want to find coefficients which have contributions only from individual spins at a given mass level. To do so for general spin and mass level, we need to consider Wilson coefficients with different mass scalings, similar to what we described with \reef{rescaleakq}.

In this spirit, we define level-$(r)$ coefficients,
\begin{align}\label{Lcoeffs}
a_{k,q}^{(r)} &\equiv \left(\ti{\mu}_r\right)^{k+1}a_{k,q} \,.
\end{align}
and their barred versions
\begin{align}
\bar{a}_{k,q}^{(r)} &\equiv \frac{a_{k,q}^{(r)}}{a_{r+\ti{J}_1,\Jt}^{(r)}}\label{Jbarred}\,.
\end{align}
We can relate barred coefficients of different level-$(r)$ by
\begin{align}\label{Lbarrel}
    \bar{a}_{k,q}^{(r_1)} = \frac{\ti{\mu}_{r_1}^{k+1}}{\ti{\mu}_{r_2}^{k+1}}\frac{a_{r_2+\ti{J}_1,\Jt}^{(r_2)}}{a_{r_1+\ti{J}_1,\Jt}^{(r_1)}}\bar{a}_{k,q}^{(r_2)}\, .
\end{align}
We chose the level-$(r)$ coefficients so that we can pick out individual tree-level contributions in the $k\to\infty$ limit, where the barred coefficients behave, for fixed $q$, as
\begin{align}\label{Jbarredscale}
\lim_{k\to\infty} \bar{a}_{k,q}^{(r)} = 
\begin{cases}
\infty & \ti{J}_1\leq q\leq r-1\\
|\bar{g}_{r,\mu_r}^{(r)}|^2v_{r,r} & q = r\\
0 & \text{otherwise} 
\end{cases}\, .
\end{align}
Crucially, crossing tells us $a_{r+\ti{J}_1,\Jt}^{(r)} = a_{r+\ti{J}_1,r}^{(r)}  \supset |g_{r,\mu_r}^{(J)}|^2v_{r,r}$, so $a_{r+\ti{J}_1,\Jt}^{(r)} \geq  |g_{r,\mu_r}^{(J)}|^2v_{r,r}$. As long as $r$ and $\mu_r$ are finite, this means
\begin{align}\label{glevbarred}
|\bar{g}_{r,\mu_r}^{(r)}|^2 \equiv \frac{|g_{r,\mu_r}|^2}{a_{r+\ti{J}_1,\Jt}^{(r)}}
\end{align}
is finite, independent of $k$, and its ratios with other level-$(r)$ barred terms can only be zero when $|\bar{g}_{r,\mu_r}^{(r)}|^2$ itself is zero.

Now, given the scale $\ti{\mu}_J$ at which spin-$J$ exchanges first appear, we can find a bound on the lowest scale $\ti{\mu}_{J+n}$ at which a spin-$J+n$ particle must appear by using Hankel matrix bounds on these level coefficients. We show that the Hankel matrix bounds require that the coupling to the spin-$J$ state be zero if $\ti{\mu}_{J+n}$ is too large as compared to $\ti{\mu}_{J}$, which causes a contradiction because $\ti{\mu}_{J}$ is defined such that there is a nonzero coupling to a spin-$J$ particle. 

The first step in finding this bound is to find a Hankel matrix bound which relates a coefficient with $q = J$, which gains contributions only from states with at least spin-$J$, to a coefficient with $q = J+n$. In the infinite $k$ limit, however, both of these coefficients vanish, so we naively get only a trivial bound by using this constraint. Therefore, in the second part of the argument, we parameterize the vanishing of these Wilson coefficients at large $k$ with the masses $\ti{\mu}_J$ and $\ti{\mu}_{J+n}$ where they respectively first have contributions. Once we have done so, we are left with an upper bound on a quantity we assume is greater than zero in terms of the $k \to \infty$ limit of a quantity which depends on $\ti{\mu}_J$, $\ti{\mu}_{J+n}$, and $\ti{\mu}_{\Jt}$, which is therefore forced to be non-vanishing. We find the upper bound on $\ti{\mu}_{J+n}$ by requiring the limit of this product of masses not go to zero.

As discussed, the first step requires the Hankel matrices. In Section \ref{sec:remsimp}, we explained that we have access only to coefficients with $q,k-q \geq \Jt$, so the Hankel matrix constraints reduce to totally nonnegativity of the submatrix
\begin{align}
\begin{pmatrix}
a_{q+\ti{J}_{1},q} & a_{q+\ti{J}_{1}+1,q} & \ldots\\
a_{q+\ti{J}_{1}+1,q} & a_{q+\ti{J}_{1}+2,q} & \ldots \\
\vdots & \vdots & \ddots
\end{pmatrix} \, .
\end{align}
By construction, these coefficients all have $q,k-q \geq \ti{J}_{1}$, so have convergent dispersion relations. The equivalent of \reef{hankbd} then is
\begin{equation}\label{hankbdn}
\begin{split}
\left(\frac{a_{k',q}}{a_{q+\ti{J}_{1},q}}\right)^{(k''-q-\ti{J}_{1})/(k'-q-\ti{J}_{1})} \leq \left(\frac{a_{k'',q}}{a_{q+\ti{J}_{1},q}}\right) \, 
\end{split}
\end{equation}
for $k\geq \ti{J}$ and $q\leq k-\ti{J}_{N}$. Taking $k' = k+J$, $k'' = k+J+n$ and $q = k$, we can rewrite this bound in terms of barred level-$(\Jt)$ coefficients as 
\begin{align}\label{hankjtbar}
    \left(\frac{\bar{a}_{k+J,J}^{(\Jt)}}{\bar{a}_{k+\Jt,\Jt}^{(\Jt)}}\right)^{\Jd_n} \leq \frac{\bar{a}_{k+J+n,J+n}^{(\Jt)}}{{\bar{a}_{k+\Jt,\Jt}^{(\Jt)}}} \, ,
\end{align}
where we define
\begin{align}
    \Jd_n = \frac{J+n-\Jt}{J-\Jt} \,.
\end{align}
This is the bound on the coefficient with $q = J$ in terms of the coefficient with $q = J+n$ we need to complete the first step of the proof.

The problem is that in the $k \to \infty$ limit, this entire expression vanishes, so gives the trivial bound $0\leq 0$. We now want to find how the Wilson coefficients vanish in terms of a finite quantity and the mass at which states first contribute to it. Using the logic of \reef{HEintbd}, we can find that for the $q = J+n$ coefficient, we have
\begin{align} \label{aJnfinbd}
    \bar{a}_{k+J+n,J+n}^{(\Jt)} \leq \frac{\ti{\mu}_{\Jt}^{k-\Jt}}{\ti{\mu}_{J+n}^{k-\Jt}}\bar{a}_{J+\Jt+n,J+n}^{(\Jt)} = \frac{\ti{\mu}_{\Jt}^{k-\Jt}}{\ti{\mu}_{J+n}^{k-\Jt}}\bar{a}_{J+\Jt+n,\Jt}^{(\Jt)} \, .
\end{align}
The basic bound \reef{basicbd} tells us $\bar{a}_{J+\Jt+n,\Jt}^{(\Jt)}$ is finite and nonzero if the theory has any contributions from spins $J +n$ or larger, which we assume it does. In the infinite $k$ limit, then, this expression vanishes not because the Wilson coefficient does, but because the SSC requires $\ti{\mu}_{\Jt} < \ti{\mu}_{J+n}$. For the $q = J$ coefficient on the other hand, \reef{Jbarredscale} tells us that
\begin{align}\label{Jbarredscalepf}
\lim_{k\to\infty} \bar{a}_{k+J,J}^{(J)} =  |\bar{g}_{J,\mu_J}^{(J)}|^2v_{J,J}\, .
\end{align}
Therefore, we want to rescale \reef{hankjtbar} so that we are bounding $a_{k+J,J}^{(J)}$ which is finite and nonzero in the large $k$ limit instead of $a_{k+J,J}^{(\Jt)}$. Doing such a rescaling and using \reef{aJnfinbd}, the Hankel bound \reef{hankjtbar} becomes
\begin{align}\label{jbarresc}
    \left(\frac{\ti{\mu}_{J}^{k+1}}{\ti{\mu}_{\Jt}^{k+1}}\frac{\bar{a}_{k+J,J}^{(\Jt)}}{\bar{a}_{k+\Jt,\Jt}^{(\Jt)}}\right)^{\Jd_n} \leq \left(\frac{\ti{\mu}_{J}^{k+1}}{\ti{\mu}_{\Jt}^{k+1}}\right)^{\Jd_n}\frac{\ti{\mu}_{\Jt}^{k-\Jt}}{\ti{\mu}_{J+n}^{k-\Jt}} \frac{\bar{a}_{J+\Jt+1,\Jt}^{(\Jt)}}{{\bar{a}_{k+\Jt,\Jt}^{(\Jt)}}} \, .
\end{align}
Using \reef{Lbarrel}, we can rewrite \reef{jbarresc} as
\begin{align}\label{mhanknolim}
    \left(\frac{a_{J+\Jt,\Jt}^{(J)}}{a_{2\Jt,\Jt}^{(\Jt)}}\frac{\bar{a}_{k+J,J}^{(J)}}{\bar{a}_{k+\Jt,\Jt}^{(\Jt)}}\right)^{\Jd_n} \leq \left(\frac{\ti{\mu}_{J}^{k+1}}{\ti{\mu}_{\Jt}^{k+1}}\right)^{\Jd_n}\frac{\ti{\mu}_{\Jt}^{k-\Jt}}{\ti{\mu}_{J+n}^{k-\Jt}} \frac{\bar{a}_{J+\Jt+1,\Jt}^{(\Jt)}}{{\bar{a}_{k+\Jt,\Jt}^{(\Jt)}}} \, 
\end{align}
so that we are explicitly bounding $a_{J+\Jt,\Jt}^{(J)}$ as desired. The $k$-independent term in the parenthesis on the left can be translated back to standard Wilson coefficients with \reef{Lcoeffs}
\begin{align}
    \frac{a_{J+\Jt,\Jt}^{(J)}}{a_{2\Jt,\Jt}^{(\Jt)}} = \frac{\ti{\mu}_J^{J+\Jt}}{\ti{\mu}_{\Jt}^{2\Jt}}\frac{a_{J+\Jt,\Jt}}{a_{2\Jt,\Jt}} \,
\end{align}
which is finite for finite $J$. The two coefficients receive contributions from all states with spins $\geq \Jt$, so there is no way their coefficient can be zero so long as there are contributions from exchanges with $J\geq \Jt$, which we are assuming is the case. Notice that we have now put all of the $k$ dependence of our bound into the ratios $\ti{\mu}_J/\ti{\mu}_{\Jt}$ and $\ti{\mu}_{J+n}/\ti{\mu}_{\Jt}$, completing the second step of our proof.

We can finally take the $k\to\infty$ limit of the Hankel bound \reef{mhanknolim}. Using \reef{Jbarredscalepf}, we see
\begin{align}
    \left(\frac{a_{J+\Jt,\Jt}^{(J)}}{a_{2\Jt,\Jt}^{(\Jt)}}\frac{|\bar{g}_{J,\mu_J}^{(J)}|^2v_{J,J}}{|\bar{g}_{\Jt,\mu_{\Jt}}^{(\Jt)}|^2v_{\Jt,\Jt}}\right)^{\Jd_n} \leq \frac{\bar{a}_{J+\Jt+1,\Jt}^{(\Jt)}}{|\bar{g}_{\Jt,\mu_{\Jt}}^{(\Jt)}|^2v_{\Jt,\Jt}} \lim_{k\to\infty} \left[\left(\frac{\ti{\mu}_{J}^{k+1}}{\ti{\mu}_{\Jt}^{k+1}}\right)^{\Jd_n}\frac{\ti{\mu}_{\Jt}^{k-\Jt}}{\ti{\mu}_{J+n}^{k-\Jt}}\right] \, .
\end{align}
As we discussed, the left hand side cannot be zero in order for $\ti{\mu}_J$ and $\ti{\mu}_{J+n}$ to be defined, so we have assumed $|\bar{g}_{J,\mu_J}^{(J)}|^2 > 0$ and that $\bar{a}_{J+\Jt+1,\Jt}^{(\Jt)} \geq 0$. Thus, we need 
\begin{align}
    \lim_{k\to\infty}\left(\frac{\ti{\mu}_{J}^{k+1}}{\ti{\mu}_{\Jt}^{k+1}}\right)^{\Jd_n}\frac{\ti{\mu}_{\Jt}^{k-\Jt}}{\ti{\mu}_{J+n}^{k-\Jt}} \neq 0 \,
\end{align}
to not have a contradiction. In order for the limit not to vanish, the masses must obey the bound \reef{massbdJ}
\begin{align}
    \frac{\ti{\mu}_{J+n}}{\ti{\mu}_{\Jt}} \leq \left(\frac{\ti{\mu}_J}{\ti{\mu}_{\Jt}}\right)^{(J+n-\Jt)/(J-\Jt)} \,.
\end{align}
The integers $J$ and $n$ are arbitrary as long as $J > \Jt$ and $0 \leq n \leq J_E-J$, so this bound holds for any such choices of $J$ and $n$ and completes the proof of the mass bound. As mentioned earlier, the constraint \reef{massbd} comes from a simple generalization of this argument and a relabeling of the indices.

\section{Discussion}\label{sec:disc}
We have shown that fundamental physical assumptions require weakly coupled color-ordered amplitudes of massless states that admit convergent $\Jt$-times subtracted dispersion relations to have states of at most spin $J = N-1+\Jt$ at the $N$th mass level for which massive states are exchanged. Additionally, the maximal spin at any given mass level can at most increase by one from the maximal spin exchanged at a lower mass level, so the existence of massive spin-$J$ exchanges necessitates exchanges of all spins lower than $J$ at some lower mass. Finally, we proved that if a state with spin-$J \geq \Jt+1$ exists, there is a strict upper bound its mass in terms of the lightest masses that spin $J-n_1$ and $J - n_2$ states appear for any $n_1,n_2 > 1$.

An additional feature important for many S-matrices with large-spin massive states is ``Reggeization'', meaning that the state with the largest mass at mass level $\mu_0$ (taken to be $\mu_0 = 0$) in this paper is a part of the leading Regge trajectory. Reggeized massless amplitudes have $n_0 \leq J_{0}$, so have convergent dispersion relations for any $k-q > J_0$. Further, their behavior in the Regge limit is controlled by the function $j(M^2)$:
\begin{align}\label{RegBeh}
    \lim_{|s|\to\infty}A(s,u) = f(u)s^{j(u)} \, ,
\end{align}
where $j(M^2)$ is defined such that $J(M^2) = j(M^2)$ on the leading Regge trajectory. Amplitudes that have states which Reggeize can be analytically continued to part of the unphysical regime in Mandelstam variables \cite{Mandelstam:1964tvk,Abers:1967zz,Gribov:2003nw}, so they exhibit the behavior \reef{RegBeh} for some $u > 0$. Based on this, the authors of \cite{Caron-Huot:2016icg} argued for a maximal spin constraint similar to \reef{spinNconst} for Reggeized scalar amplitudes. While the heuristic arguments in favor of the Froissart bound \reef{FMbound} for massless scattering typically depend on the behavior of the Legendre polynomials for $u > 0$ \cite{Arkani-Hamed:2020blm}, they still have $u \ll M_{\gap}^2$. It is quite interesting that we recover the existence of a maximal spin at a given mass level of amplitudes, which, in principle, is information contained in the leading Regge trajectory $j(u)$ for unphysical kinematics with $u > M_{\gap}^2$, where we have little analytic control over the amplitude.

Recent work with the numerical bootstrap showed that minimal input about the low-energy spectrum of the UV completion generates dramatic new features of the allowed space, including novel cusps and even reductions to shrinking islands of allowed parameter space \cite{Albert:2023bml,Berman:2024wyt,Albert:2024yap}. Upon extracting the numerically-determined high-energy spectrum of the extremal theories in these features, one finds a large number of states with low mass but high spin \cite{Caron-Huot:2021rmr,Chiang:2023quf,Albert:2023bml,Berman:2024wyt,Albert:2024yap}. The states found in the spectra of extremal theories do not appear to follow any kind of particular trajectory at all and violate the maximal spin bound we have described here. 
The results of this paper show that these states must be numerical artifacts which come from applying bounds on the truncated Lagrangian.

The SSC is only rigorous in the large $k$ limit, and so imposing it in the numerical bootstrap can be interpreted as inputting some information we know must be true in the $k_{\max} \to \infty$ limit at finite $k_{\max}$. In \cite{Haring:2023zwu,Berman:2024wyt,Albert:2024yap} this kind of constraint was input by specifying the form of the leading Regge trajectory. Rather than input a specific Regge trajectory, it would be preferable to require simply the SSC, and find the Regge trajectories which saturate the bounds. It is argued in \cite{Berman:2024wyt} that these extremal trajectories should be linear, similar to string theory, at least for some large swath of parameter space.

The mass bound is equally interesting from a bootstrap perspective. Its implementation in the numerical bootstrap procedure could also remove parameter space that cannot be ruled out at finite $k_{\max}$. Beyond the bootstrap, it also could be of some phenomenological relevance, as seen from the rather strict constraint it appears to give on the $\rho_7$ meson mass. We can continue using the measured meson masses to constrain those with even higher spin. The strongest bounds on the next few pion masses come from the upper bound $m_{f_{6}}^{J-5}/m_{\rho_5}^{J-6}$, making
\begin{equation}
\begin{split}
m_{f_8} \lesssim 2776 \text{ MeV,} \quad m_{\rho_9} \lesssim 2953 \text{ MeV,} \quad m_{f_{10}} \lesssim 3119 \text{ MeV.}
\end{split}
\end{equation}
Additionally, the mass bound appears to be applicable even beyond the limit in which it is formally derived, at least in some approximate sense. It is plausible that they could additionally be applied to any crossing symmetric system with massive states of large spin.

An obvious alternative scenario there could be an analogous constraint is for fully permutation symmetric amplitudes which are invariant under $s\leftrightarrow t \leftrightarrow u$. These bounds would be applicable to UV-completions of gravity. This scenario, however, is far more complex because there are $t$-channel poles. When we define a dispersion relation for the Wilson coefficients in these amplitudes, we need to encircle cuts on both the positive and negative $s$-axis, corresponding to the existence of $s$ and $t$-channel contributions. The basic dispersion relations become
\begin{align}
a_{k,q} = \la y^{-k}w_{J;k,q}\ra \, ,
\end{align}
where the $w_{J;k,q}$ are related, but not equal, to the Gegenbauer expansion coefficients $v_{J,q}$.
There are three main reasons the argument in Section \ref{sec:aarg} fails: the $k$ dependence of $w_{J;k,q}$, the fact that $w_{J;k,q}$ is not positive semidefinite, and the lack of $w_{J;k,q}$ coefficients that are manifestly independent of all spins up to some $J$. All of these properties prevent essential pieces of our argument from working. However, based on the stringy examples of UV completions, we still expect a type of SSC to exist. There are no contributions from odd spin exchanges to these amplitudes due to properties of the Gegenbauer polynomials, meaning the spin sum is only over even spins. Studying example amplitudes, we expect that there should be no spins above $2(N-1+\lceil{\ti{J}_1/2}\rceil)$ at the $N$th mass level and the spins should increase by at most two at each mass level.\footnote{The closed string Virasoro-Schapiro amplitude has $n_0=2$, so seems like it would be required to have at most a spin-2 at the first mass level, but it in fact has a spin-4 exchange. This is a manifestation of the graviton massless pole making $J_{0} = 2$ and $\ti{J}_1 = 3$, leading to the massive spin-4 at the first mass level.} While we do not have an analytic argument for such a bound in general, we, along with collaborators, have found numerical evidence similar to that shown in Figure \ref{fig:spincoup0} for it. 

Recently, there has been interest in writing down and checking the unitarity of new examples of tree-level UV-complete amplitudes \cite{Figueroa:2022onw,Huang:2022mdb,Geiser:2022icl,Cheung:2022mkw,Geiser:2022exp,Cheung:2023adk,Cheung:2023uwn,Haring:2023zwu,Saha:2024qpt,Wang:2024wcc,Cheung:2024uhn,Bhardwaj:2024klc,Bjerrum-Bohr:2024wyw,Albert:2024yap,Cheung:2024obl,Mansfield:2024wjc,Wang:2024jhc}. These are often generalizations of a string amplitude, whether it be Veneziano or Virasoro-Schapiro, and obey the sequential spin constraint. Our bounds can provide an explicit check of the validity of these amplitudes in addition to the simple partial-wave unitarity tests that are typically done in these papers, particularly when one cannot explicitly prove or disprove non-negativity of the partial waves. These bounds also contain more information than partial wave unitarity, as they test the explicit location of the masses and spins on a theory's leading Regge trajectory, telling us that not every function with a positive partial wave expansion is a unitary amplitude.

Despite their clear power in bounding the space of possible theories, let us emphasize that the constraints we use in this paper are not even close to the full set implied by the treatment of bounding Wilson coefficients as a moment problem. In deriving \reef{hankbd}, we use only the nonnegativity of the two-by-two minors in \reef{hankmat}, so we disregard an infinite set of inequalities implied by the total nonnegativity of \reef{hankmat}. One might hope that the remaining minors could be used to find, for example, an even stronger bound on the locations of the masses of states on the leading Regge trajectory than \reef{massbd} or a bound on the relative couplings of these states. Even beyond those larger minors, \cite{Chiang:2021ziz} showed that total nonnegativity of \reef{hankmat}, while a necessary condition, is not even sufficient to describe the full set of Hankel-type constraints on Wilson coefficients! Instead, the bounding of Wilson coefficients can be treated as a double-moment problem, leading to even more constraints. It is clear that these Hankel matrices contain important physical information, so fully understanding their implications could lead to far stronger results than the ones derived here. For example, another issue apparent in the spectra of theories in \cite{Albert:2023bml,Berman:2024wyt,Albert:2024yap} is the lack of daughter trajectories in extremal amplitudes. Similar to the existence of states above the leading trajectory, this could be a finite $k$ effect. Rather general arguments \cite{Eckner:2024pqt} rule out the existence of Reggeizing theories with single Regge trajectories, so for the extremal amplitudes to not have daughter trajectories, these amplitudes would have to be pathological when analytically continued in spin $J$, which seems unlikely due to their close connection with physical theories like real-world pion scattering and maximally supersymmetric string theory.

\section*{Acknowledgements}
I would like to thank Henriette Elvang and Nick Geiser for extensive comments and discussions as well as Aidan Herderschee and Loki Lin for additional feedback. I am also grateful for the hospitality of the Niels Bohr International Academy this fall. I am supported in part by the Cottrell SEED Award number CS-SEED-2023-004 from the Research Corporation for Science Advancement and a Leinweber Summer Fellowship. This research was supported through computational resources and services provided by Advanced Research Computing at the University of Michigan, Ann Arbor.

\appendix

\section{Full Proof for Higher Mass Levels}\label{app:hmasslev}
In this appendix, we complete the proof of the SSC with $\Jt = 0$ which we described heuristically in Section \ref{sec:hmass}. In Section \ref{sec:atmgap}, we proved that there are only scalar exchanges at the mass gap. This becomes the base $\Jt$ case for the inductive assumption that the maximal spin at mass level $n$, $J_n$, obeys the bound
\begin{align}
J_n \leq \ti{J}_{n} \, ,
\end{align}
where
\begin{align}
\ti{J}_{n} = \max\{-1,J_1 = 0,J_2,\ldots,J_{n-1}\} + 1 \,
\end{align}
is one larger than the maximal spin exchanged at any level below $n$.The goal is to show that then $J_{N} \leq \ti{J}_{N}$ if this bound holds up to some mass level $N-1$.

The $N$th mass level has $M^2 = \mu_{N}M_\gap^2$. If $\mu_N$ is not the first mass at which a state of spin $J_N$ appears, then $J_N \leq J_n$ for some $n < N$ and so $J_N < \ti{J}_N$ because $\ti{J}_N > J_n$ for any $n < N$. Therefore, the only nontrivial case to check is that in which $J_N > J_n$ for all $n < N$. This means we can assume 
\begin{align}\label{muNmutJN}
\mu_N = \ti{\mu}_{J_N} \, ,
\end{align}
where, as in Section \ref{sec:NRegTrajs}, $\ti{\mu}_{J}M_{\gap}^2$ is the lowest squared-mass at which a spin $J$ state is exchanged.

We can then use the level-$(r)$ coefficients defined in Section \ref{sec:NRegTrajs}:
\begin{align}\label{rescale1}
    a_{k,q}^{(r)} = (\ti{\mu}_{r})^{k+1} a_{k,q}  \, ,
\end{align}
with $r = J_N$. The contribution of a single tree-level exchange at the $n$th mass $M^2 = \mu_n M_{\gap}^2$ to the level-$(N)$ coefficients become
\begin{align} \label{WCtreecont1}
a_{k,q}^{(J_N)} \supset \frac{\mu_{N}^{k+1}|g_{J,\mu_n}|^2}{\mu_n^{k+1}}v_{J,q} \, ,
\end{align}
where we have used \reef{muNmutJN} to write the contributions in terms of $\mu_N$ for easier comparison with $\mu_n$. States now contribute to $a_{k,q}$ with powers of $(\mu_{N}/\mu_n)^{k+1}$, so exchanges at levels $n < {N}$ which have $\mu_n<\mu_{N}$ give problematic infinities in the large $k$ limit. However, our inductive assumption tells us that at these mass levels $n < {N}$, we can only exchange states up to some finite spin $J = \ti{J}_{N}-1$ . Since $v_{J,q} = 0$ for $J < q$, any Wilson coefficient $a_{k,q}$ with $q, k-q \geq \ti{J}_{N}$ is independent of spins $J \leq \ti{J}_{N}$, meaning they cannot have any contribution from states at mass levels $N$ or below. Therefore, any such coefficient has only finite contributions in the $k\to\infty$ limit, and we can restrict ourselves to consider only constraints on this subset of coefficients.\footnote{This is similar to considering only dispersion relations with higher levels of subtraction, as described by \reef{disprepsub}.}

We now follow the same two step process as in Section \ref{sec:atmgap}. First, we want to find a coefficient which has only contributions from the state with maximal spin $J_{N}$ at mass level $N$ in the large-$k$ limit similar to \reef{singcouplim}. Then, we need use the Hankel constraints to show that coupling gets bounded from above by zero if $J_N \geq \ti{J}_{N}$. 

The necessary coefficient can be found in a similar way to Section \ref{sec:atmgap}. Crossing symmetry, \reef{akqcrossing}, gives
\begin{equation}\label{crossedNcoeffs}
\begin{split}
&a_{k,k-J_{N}}^{(J_N)} = a_{k,J_{N}}^{(J_N)}\\
&a_{k-J_{N}+\ti{J}_{N},k-J_{N}}^{(J_N)} = a_{k-J_{N}+\ti{J}_{N},\ti{J}_{N}}^{(J_N)}\, .
\end{split}
\end{equation}
Our inductive assumption was that there are no couplings to spins $J > \ti{J}_{N}-1$ for mass levels below $N$, so there are no contributions to $a_{k-J_{N}+\ti{J}_{N},\ti{J}_{N}}$ or $a_{k,J_{N}}$ from states with mass less than $\mu_{N} M_{\gap}^2$. The dispersive expressions for the right hand sides of \reef{crossedNcoeffs} are
\begin{equation}\label{kmjmcross}
\begin{split}
&a_{k,k-J_{N}}^{(J_N)} = |g_{J_{N},\mu_{N}}|^2v_{J_{N},J_{N}} + \mu_N^{k+1}\la y^{-k}v_{J,J_{N}}\ra_{\mu_c}
\end{split}
\end{equation}
and
\begin{equation}\label{kmjmpncross}
\begin{split}
&a_{k-J_{N}+\ti{J}_{N},k-J_{N}} = \sum_{J = \ti{J}_{N}}^{J_{N}}|g_{J,\mu_{N}}|^2v_{J,\ti{J}_{N}} + \mu_N^{k+1}\la y^{-k+J_{N}-\ti{J}_{N}}v_{J,J_{N}}\ra_{\mu_c} \, .
\end{split}
\end{equation}
We now define a double indexed the level-$(r)$ barred notation as:
\begin{equation}\label{genbar}
\begin{split}
&\bar{a}_{k,q}^{(J_N,J_N)} \equiv a_{k,q}^{(J_N)}/a_{J_{N}+\ti{J}_{N},\ti{J}_{N}}^{(J_N)}\\
&|\bar{g}_{J,M^2}^{(J_N,J_N)}|^2 \equiv |{g}_{J,M^2}|^2/a_{J_{N}+\ti{J}_{N},\ti{J}_{N}}^{(J_N)} \, .
\end{split}
\end{equation}
This differs from the bar in \reef{Lbarrel} because the coefficient we normalize by is the $a_{r+\ti{J}_N,\ti{J}_N}^{(r)}$ coefficient rather than the $a_{r+\ti{J}_1,\ti{J}_1}^{(r)}$ coefficient. In this notation, then, the barred coefficients in Section \ref{sec:NRegTrajs} are the $(r,\Jt)$ coefficients.

We divide both sides of \reef{kmjmpncross} by $a_{J_{N}+\ti{J}_{N},\ti{J}_{N}}$. The naive bound \reef{basicbd} requires $\bar{a}_{k-J_{N}+\ti{J}_{N},\ti{J}_{N}}^{(J_N,J_N)}$ be bounded from above by one with $k \geq J_{N}+\ti{J}_{N}$. By the same argument as in Section \ref{sec:atmgap}, then, the high energy integral contribution to $\bar{a}^{(J_N,J_N)}_{k-J_{N}+\ti{J}_{N},\ti{J}_{N}}$ in \reef{kmjmpncross} must vanish at large $k$ to make
\begin{align}\label{qntocopl}
\lim_{k\to\infty}\bar{a}^{(J_N,J_N)}_{k-J_{N}+\ti{J}_{N},\ti{J}_{N}} = \sum_{J = \ti{J}_{N}}^{J_{N}}|\bar{g}^{(J_N,J_N)}_{J,\mu_{N}}|^2v_{J,\ti{J}_{N}} \,.
\end{align}
We chose $J_{N}+\ti{J}_{N} = J_{N}+\ti{J}_{N}$ so that, by crossing, we would have $a_{J_{N}+\ti{J}_{N},\ti{J}_{N}}^{(J_N)} = a_{J_{N}+\ti{J}_{N},J_{N}}^{(J_N)}$. The simple bound \reef{basicbd} implies $\bar{a}_{k,J_{N}}^{(J_N)}$ is also bounded from below by $\bar{a}_{J_{N}+\ti{J}_{N},\ti{J}_{N}}^{(J_N)}$ at finite $k$ and 
\begin{align}\label{qjmaxcoup}
\lim_{k\to\infty}\bar{a}_{k,J_{N}}^{(J_N,J_N)} = |\bar{g}^{(J_N,J_N)}_{J_{N},\mu_{N}}|^2v_{J_{N},\ti{J}_{N}} \, .
\end{align}
We have found in $\bar{a}^{(J_N,J_N)}_{k,k-J_{N}}$ a coefficient which depends solely on the coupling to the maximal spin-$J_N$ state.

We can use \reef{hankbdn} with $\Jt \to \ti{J}_N$ to find
\begin{equation}\label{spechankbd2}
\begin{split}
\left(\frac{a_{k,k-J_{N}}^{(J_N)}}{a_{k-J_{N}+\ti{J}_{N},k-J_{N}}^{(J_N)}}\right)^{(J_{N}-\ti{J}_{N}+1)/(J_{N}-\ti{J}_{N})} \leq \left(\frac{a_{k+1,k-J_{N}}^{(J_N)}}{a_{k-J_{N}+\ti{J}_{N},k-J_{N}}^{(J_N)}}\right) \, .
\end{split}
\end{equation}
This bound only makes sense with the hierarchy $k \geq J_{N} \geq \ti{J}_{N}$ so is not valid if $J_{N}$, the maximal spin at mass level $N$ is less than $\ti{J}_{N}$. However, we already assumed this was the case because if $J_N <\ti{J}_{N}$, \reef{maxJn} is trivially satisfied. 

Finally, we want to show that the coupling to the spin-$J_N$ state is forced to be zero for $J_{N} > \ti{J}_{N}$ by barring all coefficients in the $k\to\infty$ limit of our Hankel constraint \reef{spechankbd2} and using crossing symmetry along with \reef{kmjmcross}-\reef{qjmaxcoup}:
\begin{equation}
\begin{split}
\Bigg(\frac{|\bar{g}_{J_{N},\mu_{N}}^{(J_N,J_N)}|^2v_{J_{N},J_{N}}}{\sum_{J = \ti{J}_{N}}^{J_{N}}|\bar{g}_{J,\mu_{N}}^{(J_N,J_N)}|^2v_{J,\ti{J}_{N}}}\Bigg)^{(J_{N}-\ti{J}_{N}+1)/(J_{N}-\ti{J}_{N})} \leq \left(\frac{\lim_{k\to\infty}\bar{a}^{(J_N,J_N)}_{k+1,J_{N}+1}}{\sum_{J = \ti{J}_{N}}^{J_{N}}|\bar{g}^{(J_N,J_N)}_{J,\mu_{N}}|^2v_{J,\ti{J}_{N}}}\right) \, .
\end{split}
\end{equation}
We know $\bar{a}^{(J_N,J_N)}_{k+1,k-J_{N}} \leq \bar{a}^{(J_N,J_N)}_{J_{N}+\ti{J}_{N},k-J_{N}} = 1$ and the dispersion relation for $\bar{a}^{(J_N,J_N)}_{k+1,k-J_{N}}$ tells us
\begin{align}
\bar{a}^{(J_N,J_N)}_{k+1,J_{N}+1} = \frac{\mu_N^{k+1}\la y^{-(k+1)}v_{J,J_{N}+1}\ra_{\mu_c}}{a^{(J_N)}_{J_{N}+\ti{J}_{N},\ti{J}_{N}}} \,.
\end{align}
The high energy integral, as usual, vanishes in the infinite $k$ limit, requiring that the limit  $\lim_{k\to\infty}\bar{a}^{(J_N,J_N)}_{k+1,J_{N}+1} = 0$, meaning that $|\bar{g}^{(J_N,J_N)}_{J_{N},\mu_N}|^2 \to 0$. At mass level $N$, then, $J_{N} \leq \ti{J}_{N}$. This argument is entirely independent of $N$ other than needing $\ti{J}_{N}-1 \leq N$, and so we can minimally take $N = \ti{J}_{N}-1$, exactly where we would expect it to be if we had a string-type Regge trajectory. 

By induction, this result applies at all $N$, so crossing symmetry and the Regge behavior $\lim_{|s|\to\infty}A(s,u) \to 0$ imply that any amplitude subject to our assumptions has a maximal spin of $\ti{J}_{N}$ at mass level $N$ in its spectrum. Further, the highest spin in the massive spectrum must increase one at a time: an amplitude having exchanges of states with spin $J$ at some mass necessitates the existence of states with spin $J-1$ at some lower mass. Together, these conditions tell us that, at a given mass level $N$, the maximal allowed spin has the bound
\begin{align}\label{pvSSCapp}
    J_N \leq \ti{J}_{N} = \max\{-1,J_1,J_2,\ldots,J_{N-1}\} +1\,.
\end{align}
This holds at any $N$, telling us spins can increase at most by one at each mass level. Since we start from only spin zero states at mass level one, we know that
\begin{align}
J_N \leq N-1
\end{align}
for any mass level $N$, so we know
\begin{align}
    J_N \leq \ti{J}_{N} = \max\{-1,J_1,J_2,\ldots,J_{N-1}\} +1 \leq N-1\,.
\end{align}
matching \reef{spinNconst} with $\Jt = 0$.

\section{Deriving \reef{hankbd}} \label{app:Hankbd}
We start from the compact expression of the Wilson coefficients \reef{finalequ}, which we restate here for convenience:
\begin{equation}\label{finalequapp}
a_{k,q}\,=\,\sum_{J=0}^{\infty}\int_{1}^\infty dy \,f_{J}(y) \,y^{-k} v_{J,q} ,.
\end{equation}
This expression can be converted to a so-called ``Hausdorff moment problem'' by changing variables to $x = \frac{1}{y}$, giving 
\begin{equation}\label{SM1}
a_{k,q}\,=\,\sum_{J=0}^{\infty}\int_{0}^{1} \frac{dx}{x^2}\,x^{k+D/2+1}\text{Im}(a_{J}\big(M_{\gap}^2/x)\big) \,x^{k} v_{J,q} ,.
\end{equation}
Remembering that $a_{k,q}$ need only be convergent for $k \geq \Jt + q$, we then define the measure 
\begin{align}
d\mu_q (x) =  \frac{dx}{x^{1-D/2-\Jt-q}}\sum_{J=0}^{\infty}\text{Im}(a_{J}\big(M_{\gap}^2/x)\big)v_{J,q}\, ,
\end{align}
which is a positive Borel measure because $x \geq 0$, $\text{Im}(a_{J}(M_{\gap}^2/x)) > 0$ from unitarity, and $v_{J,q} \geq 0$ was a property given in the main text. Then, defining $m = k-q-\Jt$, \reef{SM1} can be written in the canonical form of a Hausdorff moment problem,
\begin{align}\label{hmomp}
    a_{k,q} = \int_{0}^1 x^{m} d\mu_q(x)\, ,
\end{align}
with $m \geq 0$. We know there exists a solution to this moment problem because the $a_{k,q}$ coefficients are all convergent for the chosen $m$. The existence of a solution to the Hausdorff moment problem requires that the associated Hankel matrices are positive semidefinite:
\begin{align} \label{hmat1psd}
    \begin{pmatrix}
        a_{q+\Jt,q} & a_{q+\Jt+1,q} & \ldots\\
        a_{q+\Jt+1,q} & a_{q+\Jt+2,q} & \ldots \\
        \vdots & \vdots & \ddots
    \end{pmatrix} \succeq 0
\end{align}
and 
\begin{align}
    \begin{pmatrix}
        a_{q+\Jt+1,q} & a_{q+\Jt+2,q} & \ldots\\
        a_{q+\Jt+2,q} & a_{q+\Jt+3,q} & \ldots\\
        \vdots & \vdots & \ddots
    \end{pmatrix} \succeq 0 \, .
\end{align}
This condition is equivalent to the statement that \reef{hmat1psd} be totally nonnegative \cite{Pinkus:2010tpm}. As discussed in the text, total nonnegativity means all minors of \reef{hmat1psd} must have nonnegative determinant. To derive \reef{hankbd}, we need consider only determinants of the two-by-two minors in the first two rows
\begin{align}\label{minorbd}
    0 \leq a_{q+n,q}a_{q+n+2,q} - a_{q+n+1,q}^2 \,
\end{align}
Defining
\begin{align}
    \ti{a}_{k,q} = \frac{a_{k,q}}{a_{q+\Jt,q}} \,,
\end{align}
\reef{minorbd} requires
\begin{align}\label{timinorbound}
\ti{a}_{k,q}^2 \leq \ti{a}_{k-1}\ti{a}_{k+1,q}
\end{align}
for all $k > q + \Jt$. Starting from $k = q + \Jt+1$, we find
\begin{align}
\ti{a}_{q+\Jt+1,q}^2 \leq \ti{a}_{q+\Jt+2,q} \, .
\end{align}
Assume for induction that 
\begin{align}\label{indass}
\ti{a}_{k,q}^{(k-q-\Jt+1)/(k-q-\Jt)} \leq \ti{a}_{k+1,q} \, ,
\end{align}
which is satisfied at the base case $k = q+\Jt+1$. By \reef{timinorbound}
\begin{align}
    \ti{a}_{k+1,q}^2 &\leq \ti{a}_{k}\ti{a}_{k+2} \leq \ti{a}_{k+1}^{(k-q-\Jt)/(k-q-\Jt+1)}\ti{a}_{k+2},
\end{align}
where the second inequality comes from changing the location of the power in our in \reef{indass}. Then, we can simply rearrange to find
\begin{align}
    \ti{a}_{k+1,q}^{2-(k-q-\Jt)/(k-q-\Jt+1)} = \ti{a}_{k+1,q}^{(k-q-\Jt+2)/(k-q-\Jt+1)} \leq \ti{a}_{k+2} \, ,
\end{align}
proving our inductive hypothesis. To get a relationship between general $\ti{a}_{k,q}$ and $\ti{a}_{k',q}$, simply consider

\begin{equation}
\begin{split}
     \ti{a}_{k,q}\leq \ti{a}_{k+1,q}^{(k-q-\Jt)/(k-q-\Jt+1)} &\leq \ti{a}_{k+2,q}^{(k-q-\Jt)/(k-q-\Jt+1)(k-q-\Jt+1)/(k-q-\Jt+2)} \\
     &= \ti{a}_{k',q}^{(k-q-\Jt)/(k-q-\Jt+2)} \leq \ldots \leq \ti{a}_{k,q}^{(k-q-\Jt)/(k'-q-\Jt)} \,.
\end{split}
\end{equation}
Writing these out in terms of bare Wilson coefficients and moving the exponent, we finally find
\begin{align}
    \left(\frac{a_{k,q}}{a_{q+\Jt,q}}\right)^{(k'-q-\Jt)/(k-q-\Jt)} \leq \frac{a_{k',q}}{a_{q+\Jt,q}} \,,
\end{align}
matching the bounds in \reef{hankbd} for $\Jt = 0$ and \reef{hankbdn} for generic $\Jt$.

\bibliographystyle{JHEP}
\bibliography{MaxSpin.bib}
\end{document}